\documentclass[10pt,journal,compsoc]{IEEEtran}

\ifCLASSOPTIONcompsoc
  \usepackage[nocompress]{cite}
\else
  \usepackage{cite}
\fi
\ifCLASSINFOpdf
\else
\fi
\usepackage{amsthm}
\newtheorem{defn}{\textbf{Definition}}
\newtheorem{proposition}{Proposition}

\newtheorem*{problemstate}{\textbf{Problem Statement}}
\usepackage{graphicx}
\usepackage{subfigure}
\usepackage{amsmath}
\usepackage{amssymb}
\usepackage{mathrsfs}
\usepackage[ruled, vlined, linesnumbered]{algorithm2e}
\usepackage[hidelinks]{hyperref}
\usepackage{cite}
\usepackage{makecell}
\usepackage[numbers]{natbib}
\usepackage{array}
\usepackage{tikz}
\usetikzlibrary{trees}
\usepackage{tabularx}
\usepackage{threeparttable}
\usepackage{booktabs}
\usepackage{multirow}
\usepackage{longtable}
\usepackage{rotating}
\usepackage[capitalize,noabbrev]{cleveref}
\crefname{equation}{Eq.}{Eqs.}
\usepackage{ragged2e}

\begin{document}
%
\title{CRFU: Compressive Representation Forgetting Against Privacy Leakage on Machine Unlearning}

\author{Weiqi~Wang, ~\IEEEmembership{Member,~IEEE},
	Chenhan~Zhang, ~\IEEEmembership{Member,~IEEE},
        Zhiyi~Tian, ~\IEEEmembership{Member,~IEEE},\\
        Shushu~Liu, ~\IEEEmembership{Member,~IEEE}
        and Shui~Yu,~\IEEEmembership{Fellow,~IEEE}
\IEEEcompsocitemizethanks{
	\IEEEcompsocthanksitem W. Wang, Z. Tian and S. Yu are with the School of Computer Science, University of Technology Sydney, Australia.\protect\\
	E-mail: { Weiqi.Wang-2@student.uts.edu.au, \\ \{zhiyi.tian-1, shui.yu\}@uts.edu.au}
	\IEEEcompsocthanksitem C. Zhang is with the Postdoctoral Research Fellow, Macquarie University, Australia. \protect 
	E-mail:  chzhang@ieee.org
	\IEEEcompsocthanksitem S. Liu is with the Department of Communication and Networking, Aalto University, Espoo, Finland.\protect 
	E-mail: liu.shushu@aalto.fi 
}
\thanks{This paper was supported in part by Australia ARC LP220100453, ARC DP200101374, and ARC DP240100955. \\ \it{(Corresponding author: Chenhan Zhang.)}}
}

%
%

\markboth{This paper is submitted to IEEE Transactions on Dependable and Secure Computing}%
{Shell \MakeLowercase{\textit{et al.}}: Bare Advanced Demo of IEEEtran.cls for IEEE Computer Society Journals}
%

\IEEEtitleabstractindextext{%
\begin{abstract}	
\justifying
Machine unlearning allows data owners to erase the impact of their specified data from trained models. Unfortunately, recent studies have shown that adversaries can recover the erased data, posing serious threats to user privacy. An effective unlearning method removes the information of the specified data from the trained model, resulting in different outputs for the same input before and after unlearning. Adversaries can exploit these output differences to conduct privacy leakage attacks, such as reconstruction and membership inference attacks. However, directly applying traditional defenses to unlearning leads to significant model utility degradation. In this paper, we introduce a Compressive Representation Forgetting Unlearning scheme (CRFU), designed to safeguard against privacy leakage on unlearning. CRFU achieves data erasure by minimizing the mutual information between the trained compressive representation (learned through information bottleneck theory) and the erased data, thereby maximizing the distortion of data. This ensures that the model's output contains less information that adversaries can exploit. Furthermore, we introduce a remembering constraint and an unlearning rate to balance the forgetting of erased data with the preservation of previously learned knowledge, thereby reducing accuracy degradation. Theoretical analysis demonstrates that CRFU can effectively defend against privacy leakage attacks. Our experimental results show that CRFU significantly increases the reconstruction mean square error (MSE), achieving a defense effect improvement of approximately $200\%$ against privacy reconstruction attacks with only $1.5\%$ accuracy degradation on MNIST.

\end{abstract}

}

\maketitle

\IEEEdisplaynontitleabstractindextext

%
\IEEEpeerreviewmaketitle

\ifCLASSOPTIONcompsoc
\IEEEraisesectionheading{\section{Introduction}\label{sec:introduction1}}
\else
\section{Introduction}
\label{sec:introduction}
\fi

Machine unlearning has garnered extensive attention recently, as the ``right to be forgotten'' has been legislated worldwide. This legislation mandates that Machine Learning (ML) service providers not only delete data collected from individual users but also remove the contribution of that data from trained ML models \cite{mantelero2013eu}. The concept of machine unlearning was first introduced by \citeauthor{cao2015towards} in \cite{cao2015towards}, involving the removal of a specific subset of data previously used to train an ML model. ML model providers must ensure that the provided model is no longer associated with the erased data. Various methods, including exact unlearning \cite{cao2015towards,bourtoule2021machine} and approximate unlearning \cite{nguyen2020variational,fu2022knowledge}, have been proposed as effective solutions to exercise users' rights and protect their privacy.

Although machine unlearning aims to preserve users' privacy by removing their data from trained models, recent studies have revealed that the changes induced by unlearning can leak private information about the erased data. For instance, \citeauthor{chen2021machine}~\cite{chen2021machine} proposed membership inference attacks to infer which data samples are present in the erased dataset. Additionally, \citeauthor{zhang2023conditional} and \citeauthor{Hu2024sp}~\cite{gao2022deletion,Hu2024sp,zhang2023conditional} identified a further attack, known as privacy reconstruction, capable of recovering deleted data based on unlearning updates. Similarly, \cite{salem2020updates} introduced the privacy reconstruction attack aimed at recovering specific data points used in model updates. The privacy leakage caused by unlearning updates has become a critical issue in machine unlearning, and few works have effectively addressed this threat, particularly concerning reconstruction attacks.

Addressing potential privacy breaches in machine unlearning is critical and requires urgent attention. However, finding an effective solution remains a significant challenge. Firstly, existing unlearning methods, while effective, conflict with privacy leakage defenses because unlearning necessitates the complete removal of the specified samples' information from the trained model. This typically results in different outputs for the same query set before and after unlearning, which adversaries can exploit to infer private information about the erased data. Secondly, directly incorporating defense strategies such as differential privacy \cite{dwork2006differential} into unlearning methods exacerbates model utility degradation, leading to what is termed ``unlearning catastrophic''~\cite{nguyen2020variational,du2019lifelong}, rendering the model unusable. The key challenge lies in designing an unlearning method that reduces the private information in the model output to prevent inference attacks while still achieving effective unlearning performance.

\begin{figure}[tbp]
	\centering
	\includegraphics[width=0.49\textwidth]{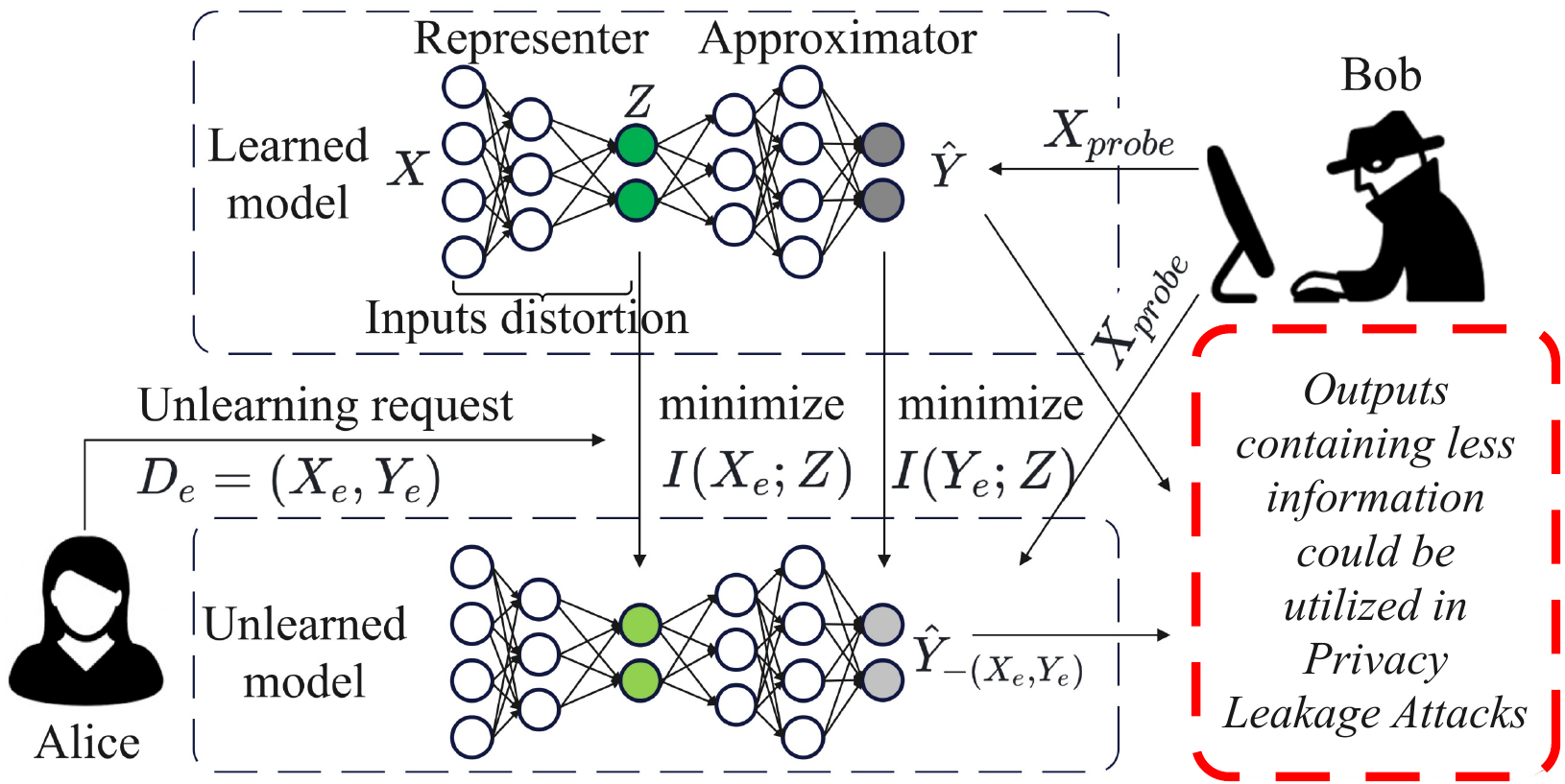}
	\caption{Compressive representation forgetting unlearning (CRFU) to defend against privacy leakage on unlearning. When a user (Alice) requests to unlearn her sensitive dataset $D_e$ from a trained IB model, CRFU minimizes the mutual information between the erased dataset and the learned representation to unlearn both the representer and approximator of this model. Since both our unlearning and original learning distorts as much information of inputs $X$ in representation $Z$ as possible, the outputs of the model contain less information about the inputs that could be used for privacy inference by Bob. It can also be explained from the training Markov chain, $X \to Z \to \hat{Y}$; and the representer is trained to minimize the mutual information between X and Z; therefore, $I(X;X) \gg I(X;Z) \ge I(X;\hat{Y})$. }
	\label{unlearning_sys}
\end{figure}

\noindent \textbf{Our work:} In this paper, we propose a Compressive Representation Forgetting Unlearning (CRFU) scheme to defend against privacy leakage on machine unlearning, as illustrated in \Cref{unlearning_sys}. The CRFU scheme is tailored for unlearning models trained using information bottleneck (IB) theory~\cite{tishby2000information}. An IB-trained model comprises two components, a representer and an approximator, which aim to maximize the compression of input data while preserving sufficient information relevant to the labels in the learned representation. CRFU implements unlearning for both components. Specifically, CRFU further minimizes the mutual information between learned representations and the erased data (both inputs and labels) to remove the information of these samples in IB-trained models. Since the unlearning process is based on representations, which maximizes the distortion of both the original training data inputs and the erased data inputs, the model's output contains less information that adversaries can exploit. Consequently, CRFU can effectively defend against privacy leakage attacks for a black-box ML model \cite{papernot2017practical}. However, directly optimizing unlearning methods would lead to significant model utility degradation, known as ``catastrophic unlearning.'' To counteract model utility decline, we introduce a remembering constraint and an unlearning rate in CRFU. Adjusting the unlearning rate helps balance the trade-off between completely removing the influence of specified samples and preserving previously acquired knowledge.

We conduct theoretical analysis and extensive experiments to demonstrate that CRFU achieves a significant improvement in defense against privacy leakage attacks~\cite{Hu2024sp,chen2021machine} than existing unlearning methods \cite{guo2019certified,nguyen2020variational}. 
Moreover, we evaluate the erasure effectiveness of CRFU employing a widely recognized methodology outlined in \cite{hu2022membership}. This involves strategically integrating backdoored samples into the training dataset used for the original model training. The primary goal of the unlearning process is to comprehensively remove the influence of these backdoored samples on the trained model. The efficacy is evaluated based on the performance of backdoor attacks: a lower attack success rate signifies superior unlearning performance. The results show that CRFU significantly outperforms the recently proposed state-of-the-art approximate unlearning methods \cite{guo2019certified,nguyen2020variational,fu2022knowledge} in both privacy leakage defense and erasure effectiveness.

To sum up, our contributions are:
\begin{itemize}
	\item To our best knowledge, this is the first work that studies the defense of privacy leakage on unlearning. Our approach, compressive representation forgetting unlearning (CRFU), can achieve a significant defense effect and is easy to extend to support most ML algorithms. 
	\item We introduce a remembering constraint and an unlearning rate in CRFU to achieve a balance between forgetting the erased samples and remembering previously learned knowledge. Adjusting the unlearning rate can control the unlearning extent and effectively mitigate the utility decline.
	\item Theoretical analysis and experimental results demonstrate the effect of CRFU in defending against privacy leakage and mitigating model utility degradation, which outperforms state-of-the-art unlearning methods. The codebase is accessible at \url{https://github.com/wwq5-code/CRFU.git}. 
\end{itemize}

The rest of the paper is structured as follows. We review the related work in Section \ref{rw}. An overview of the required background and notation can be found in Section \ref{pr}. In Section \ref{rfunl}, we first define the compressive representation forgetting unlearning problem based on IB theory; then, we present a detailed introduction to compressive representation forgetting unlearning. Section \ref{theoretical_a} provides a theoretical analysis of why CRFU can defend against privacy leakage on unlearning. Section \ref{ex} details our experimental findings, showcasing a comparative analysis with related work. In Section \ref{summary}, we offer a concise summary of the paper.

\section{Related Work}
\label{rw}

\subsection{Machine Unlearning}
 
Machine unlearning, a method designed to eliminate the contribution of certain data from a trained model, facilitates individuals' right to be forgotten. This concept is explored in works such as \cite{graves2021amnesiac,garg2020formalizing}. A straightforward but often impractical approach to machine unlearning involves retraining the machine learning model from scratch. This method can be prohibitively expensive in terms of computational overhead and storage requirements, particularly for complex deep learning tasks. Current strategies for machine unlearning are generally divided into two main types: ``fast retraining'' and ``approximate unlearning''.

Fast retraining methods in machine unlearning primarily focus on reducing the computational overhead associated with retraining models. These methods involve a partial redesign of learning algorithms and necessitate storing training data or intermediate parameters during the training process, thereby incurring increased storage costs \cite{cao2015towards,bourtoule2021machine,yanarcane2022unlearning,wu2022puma}. In \cite{cao2015towards}, \citeauthor{cao2015towards} restructured traditional machine learning algorithms into a summation-based framework. This innovation allows for quicker updates to the model when unlearning is required. Instead of retraining the entire model from scratch, only a few summations need to be modified, significantly accelerating the process. In \cite{bourtoule2021machine,yanarcane2022unlearning}, \citeauthor{bourtoule2021machine} and \citeauthor{yanarcane2022unlearning} proposed advanced methods for unlearning samples in deep neural networks. Their approach involves segmenting the complete dataset into smaller portions, referred to as shards, with individual sub-models trained on each shard. When data needs to be unlearned, it is removed from the relevant shard, and only the sub-model for that specific shard is retrained. This method, however, incurs significant storage costs, as it requires maintaining the intermediate training parameters for each shard and storing the entire training dataset. Additionally, its efficiency diminishes with an increase in the frequency of data removal requests.

Approximate unlearning methods aim to directly modify the original trained model to approximate ground truth (model retrained using the remaining dataset).
In \cite{guo2019certified,sekhari2021remember}, \citeauthor{guo2019certified} introduced a certified-removal method that draws parallels to the principles of differential privacy, as established by \citeauthor{dwork2006differential}\cite{dwork2006differential}. The mechanism ensures that a model, after data erasure, remains indistinguishable from a model that never incorporated the erased data. To address the potential decline in model utility due to unlearning, both \cite{guo2019certified} and \cite{sekhari2021remember} employed a strategy that limits the updates to model parameters. This approach aligns with the concept of differential unlearning, maintaining model effectiveness while adhering to unlearning requirements. In \cite{nguyen2020variational}, \citeauthor{nguyen2020variational} approached unlearning by approximating the posterior based on the data set for removal, utilizing Bayesian inference techniques as outlined in \cite{box2011bayesian}. While these methods employ bounds or thresholds to avert significant loss of unlearning, they can still lead to a reduction in model utility after unlearning. To address this, we introduce a remembering constraint and an unlearning rate in CRFU to diminish the accuracy decline caused by unlearning.



\subsection{Privacy Attacks on Unlearning}
As machine unlearning becomes hot, it also brings new challenges of privacy threats. \citeauthor{chen2021machine}~\cite{chen2021machine} highlighted that the variance in a model's outputs before and after unlearning could inadvertently expose the privacy of the removed data. They further explored this issue by proposing a membership inference attack specifically targeting the unlearning process. \citeauthor{lu2022label} \cite{lu2022label} further proposed label‐only membership inference attacks targetting black-box machine unlearning. In addition to membership inference attacks on unlearning, \citeauthor{gao2022deletion}~\cite{gao2022deletion} introduced the concept of deleted reconstruction attacks. This form of model inversion attack~\cite{fredrikson2014privacy}, aims to reconstruct removed data by analyzing the outputs of both the original and the unlearned models. Similarly, \citeauthor{salem2020updates}~\cite{salem2020updates} proposed a reconstruction attack aimed at recovering specific data samples involved in the model updating process. This is achieved by comparing the model's outputs before and after the update.

However, there are few works have effectively addressed the threat caused by machine unlearning. In \cite{chen2021machine}, \citeauthor{chen2021machine} suggested various strategies to mitigate privacy leaks in machine unlearning, one of which includes the application of differential privacy. However, they noted that while these methods can enhance privacy protection, they often exacerbate the issue of utility degradation in the model. In the paper, we introduce CRFU, a novel approach that is tailored to unlearning IB-trained models by discarding the information of the specified data from the trained compressed representations. Since the data has been compressed during learning and unlearning, it can effectively thwart adversaries' attempts to infer private properties from the model outputs.

\section{Preliminary}
\label{pr}

 \begin{table}[t]
 	\renewcommand\arraystretch{1.2}
	\centering  
	\scriptsize
	\caption{Basic Notations \vspace{-2mm}}
	\label{notation}
	\resizebox{\linewidth}{!}{ \huge
		\begin{tabular}{c || l }  
			\toprule[0.12em]
			Notations &Descriptions \\ \hline  
			$D=(X,Y)$ & \makecell[l]{Full training data $D$, including inputs  $X$ and  labels $Y$}   \\     
			$D_e=(X_e,Y_e)$ &\makecell[l]{The erased data $D_e$, including inputs $X_e$ and labels $Y_e$ }   \\   
			$D_r=(X_r,Y_r)$ &\makecell[l]{The remaining data $D_r$, including $X_r$ and $Y_r$ }   \\  
			\makecell[c]{$Z$} &The compressive representation \\  
			\makecell[c]{$p(Z|X)$} &\makecell[l]{The representation posterior learned based on $X$}\\  
			\makecell[c]{$p(Z|X_{-X_e})$} &\makecell[l]{The representation posterior unlearned based on $X_e$}  \\  
			$\hat{Y}$ & The predicted approximation of the model \\  
			$\hat{Y}_{-(X_e,Y_e)}$ & \makecell[l]{ The unlearned approximation, also being written as $\hat{Y}_u$} \\  
			$x, z, y$ & The persample from $X, Z, Y$ \\  
			$\theta^r$& The representer of an IB model \\  
			$\theta^a$& The approximator of an IB model \\  
			$\mathcal{L}_{rep}$ & The representation loss\\  
			$\mathcal{L}_{app}$ & The approximation loss of predicting \\  
			$\mathcal{L}_{rep}^u$ & The representation unlearning loss\\  
			$\mathcal{L}_{app}^u$ & The approximation unlearning loss\\   
			$\beta$ & The Lagrange multiplier in IB \\  
			$\beta_u$ & \makecell[l]{The unlearning rate of CRFU } \\ 
			\bottomrule[0.12em]
	\end{tabular}}
\end{table}



In Table \ref{notation}, we present a concise summary of the key notations employed throughout this paper. 
The full training dataset is represented as $D$, consisting of the input data $X$ and labels $Y$. The notation $Z$ is used to denote the compressive representation learned through the IB model, which is optimized to maximize the distortion of input data while concurrently retaining maximal relevant information about the labels. The representation posterior, formulated from $X$, is expressed as $p(Z|X)$. Additionally, the unlearned representation posterior, unlearning based on the specified data $X_e$, is denoted as $p(Z|X_{-X_e})$. The notation $\hat{Y}$ is used to utilized to represent the predictive approximation output of an IB model. After unlearning, the predictive approximation is denoted either as $\hat{Y}_{-(X_e,Y_e)}$ or $\hat{Y}_u$. In experimental setups, we use sample variables $x, z, y$ drawn from populations $X, Z, Y$. In IB algorithms, there are two types of losses: representation loss ($\mathcal{L}_{rep}$) and prediction approximation loss ($\mathcal{L}_{app}$). Here, $\beta$ acts as a Lagrange multiplier in IB, balancing distortion and utility. In the CRFU process, we encounter the final unlearning loss, a combination of representation unlearning loss ($\mathcal{L}_{rep}^u$) and approximation unlearning loss ($\mathcal{L}_{app}^u$). The parameter $\beta_u$ is used to manage the trade-off between completely unlearning the influence of specified samples and not entirely forgetting the previously learned representation.

\subsection{Threat Model of Privacy Leakage Attacks on Unlearning}
 
One effective privacy leakage attack is called model inversion, initially introduced by \citeauthor{fredrikson2014privacy} \cite{fredrikson2014privacy}. It aims to deduce missing attributes of input features by interacting with a trained ML model. An example of this is the reconstruction of facial images from deep learning models, as studied in \cite{mai2018reconstruction}. In the context of machine unlearning, researchers like \citeauthor{Hu2024sp} \cite{gao2022deletion,Hu2024sp,chen2021machine,zhang2023conditional} have identified that the discrepancies in a model's outputs before and after unlearning could potentially reveal information about the deleted data. These attacks in the realm of unlearning primarily focus on exploiting these differences to compromise the erased data privacy. The main process and adversary capabilities of privacy leakage attacks on unlearning are summarized as follows. 

\begin{figure}[t]
	\centering
	\includegraphics[width=0.48\textwidth]{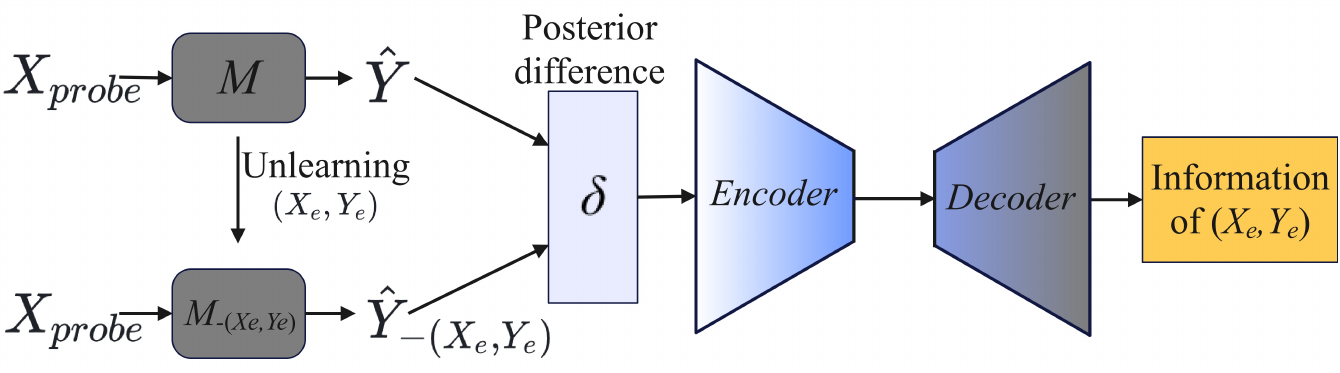}
	\vspace{-2mm}
	\caption{Privacy leakage attack based on the black-box model's outputs before and after unlearning.}
	\label{reconstruction_attack}
\end{figure}


\subsubsection{Attacking Data Preparation}
Assuming $D$ represents the original training dataset and $e$ is the sample a user requests to unlearn, we denote $D_r = D \backslash \{e\}$ as the remaining dataset. If there are several erased samples, we refer to the erased dataset as $D_e$. Machine unlearning aims to erase the influence of the specified dataset $D_e$ from a trained ML model $\mathcal{M}(D)$, where the model is trained based on the full training dataset $D$. The unlearning update is performed by executing an unlearning algorithm $\mathcal{U}$ on the current trained model using the erased dataset. More formally, given a erased dataset $D_e$, a trained ML model $\mathcal{M}$, and the unlearning algorithm $\mathcal{U}$, the unlearning process can be defined as $\mathcal{U}: D_e, \mathcal{M}(D) \to \mathcal{M}_u(D\backslash D_e)$, where $\mathcal{M}_u$ is the unlearned version model $\mathcal{M}$. 

\subsubsection{Adversary Goal and Capabilities}
The goal of privacy leakage attacks on machine unlearning is to infer the privacy of the erased samples $D_e$ of an unlearning update.
We assume the adversary can only have black-box access to the target model. 
This limitation means that the adversary can interact with the model solely through queries using a specific set of data samples, known as the probing set, and subsequently receive the corresponding outputs. Furthermore, it is presumed that the adversary's local probing dataset is sourced from the same distribution as that of the target model's training dataset. Moreover, we consider that the adversary has knowledge of both the unlearning and original learning algorithms and has the capability to establish the same learning and unlearning training as the target model. 
This can be achieved by performing model hyperparameter stealing attacks \cite{wang2018stealing}.
Finally, and most importantly, we assume that the target model effectively removes the information of the erased data, and the erased dataset and the remaining dataset are disjoint.

\subsubsection{Privacy Inference Attack Process on Unlearning}

In privacy inference attacks, the adversary's initial step involves collecting varied outputs from their probe data $X_{probe}$. This includes obtaining the original model outputs $\hat{Y}$ before unlearning, as well as the outputs $\hat{Y}_{-(X_e, Y_e)}$ after unlearning. The key to the attack lies in the difference $\delta = \hat{Y}_{-(X_e,Y_e)}- \hat{Y}$, which is used to train the attack model. The structure of the attack model typically includes an encoder and a decoder, resembling the architecture of Variational Autoencoders (VAEs) \cite{kingma2014auto}. We replicate the privacy inference attacks following the methodology outlined by \cite{salem2020updates}, and the detailed attack process is depicted in \Cref{reconstruction_attack}. The direct victims are those users who request for unlearning. Their unlearning requests prompt the ML server to execute an unlearning update, which is leveraged by this kind of attack to infer the privacy of the erased data $(X_e,Y_e)$.


\subsection{Implementation of Information Bottleneck}

Information Bottleneck (IB) was first introduced in \cite{tishby2000information} to distort information of data inputs while maintaining the information of data targets in the representation. Traditional IB~\cite{tishby2000information,tishby2015deep} primarily utilizes the Blahut-Arimoto algorithm~\cite{blahut1972computation} to optimize the IB objective. The goal of this approach is to identify a compressed encoding distribution that is both adequate for the intended ML application and maximally distorts information from the original data. The IB objective is formulated as below,
\begin{equation} \label{IB_eq}
	\mathcal{L}_{\mathcal{IB}} =   \beta I(Z;X) - I(Z;Y).
\end{equation}
Here, $I(Z; X)$ represents the mutual information between the encoded representation $Z$ and the inputs $X$, and $I(Z; Y)$ denotes the mutual information between $Z$ and the outputs $Y$. The parameter $\beta$ serves as a Lagrange multiplier, regulating the distortion ratio of $X$ in the model.


An IB model splits the training process into two parts. Firstly, it employs a representer $\theta^r$ to compress the information from inputs $X$ to a compact representation $Z$. Secondly, it employs an approximator $\theta^a$ to identify the target values based on the representation $Z$. The representation loss function of the representer is $\mathcal{L}_{rep} =  \beta I_{\theta^r}(Z;X)$ and the corresponding approximation loss function of the approximator is $\mathcal{L}_{app} = - I_{\theta^a}(Z;Y)$. The training process follows the Markov chain $Y \to X \to Z \to \hat{Y}$, as depicted in the upper part of \Cref{fig:unlvibrfu}. Prior studies, such as \cite{alemi2016deep,achille2018information,bang2021explaining}, have expanded on the mutual information terms and introduced two variational distributions, $q(Z)$ and $q(Y|Z)$. These distributions help in deriving an upper bound for the IB optimization loss function. The optimization of the learning process is then achieved by minimizing this upper bound. We consider a distribution $q(Z)$ where the elements in the space $Z$ are mutually independent, expressed as $q(Z) = \prod_j q_j(z_j)$. Finally, an IB loss objective \Cref{IB_eq} can be expanded in a per-sample way as 
\begin{equation} \label{eq:total_loss_q}
	\begin{small}
		\begin{aligned}
			\mathcal{L} = \mathcal{L}_{app} + \mathcal{L}_{rep}  = \frac{1}{N} \sum_{i=1}^{N} \mathbb{E}_{z \backsim p_{\theta^r}(Z|x^i)}[-\log \ p_{\theta^a}(y^i|z)] +\\
			+ \beta \text{KL}[p_{\theta^r}(Z|x^i)||  \prod_{j}^{|Z|} q^i_j(z^i_j)].
		\end{aligned}
	\end{small}
\end{equation}
It's important to note that the assumed prior $q(Z)$ represents the ideal distribution, capturing minimal information about $X$ while retaining sufficient information for the task target $Y$. A higher value of $\beta$ leads to increased distortion, making $p(Z|X)$ more closely approximate the assumed general prior. On the contrary, a smaller $\beta$ means the model pays less attention to compressing the information of $X$, causing the representation to contain more information about individual $X$.

\section{Compressive Representation Forgetting Unlearning}\label{rfunl}
 
\subsection{Problem Definition} \label{problem_df}

We define the problem of machine unlearning in the context of trained IB models as follows. Upon receiving an unlearning request, the full training dataset $D=(X,Y)$ will be partitioned into two distinct subsets: a smaller erased dataset $D_e=(X_e,Y_e)$ and a larger remaining dataset $D_r=(X_r, Y_r)$. 
These subsets are mutually exclusive, fulfilling $D = D_r  \bigcup D_e $ and $D_r \bigcap D_e = \emptyset$. 
Then, the IB algorithm ($\mathcal{IB}$) is used to train the model $\mathcal{M}(\theta^r, \theta^a)$, consisting of a representer $\theta^r$ and an approximator $\theta^a$. This model learns a representation $Z$, by minimizing $I(X;Z)$ and maximizing $I(Y;Z)$. 

Typically, the machine unlearning process is initiated when a user requests the removal of their specific data samples, $D_e$, from the trained model $\mathcal{M}(\theta^r, \theta^a)$. In response, the server employs an unlearning algorithm $\mathcal{U}$ to erase the contribution of $D_e$ while retaining the learned knowledge on the remaining dataset $D_r$. The unlearned model $\mathcal{M}_u(\theta^r, \theta^a)$ hopes to find the unlearned representation $p(Z|X_r)$, which is expected to be equal to the representation learned by retraining the model based on the remaining dataset. 
The problem of unlearning an IB model can be described as 
\begin{problemstate}
Assume a gold-standard unlearned representation posterior $p(Z|X_r)$ for an IB model, retrained with $\mathcal{M}_u =  \mathcal{IB}(D_r)$ based on the remaining dataset. An unlearning algorithm $\mathcal{U}$ is designed to unlearn a representation posterior $p(Z|X_{-X_e})$ of the model $\mathcal{M}_u =  \mathcal{U}(D_e, \mathcal{IB}(D))$, unlearning based on the erased dataset $D_e=(X_e, Y_e)$. Then, the erased dataset-based unlearned posterior is hoped to match the remaining dataset-based retrained posterior, i.e., 
\begin{equation} \label{eq_eu}
p(Z|X_r) = p(Z|X_{- X_e}).
\end{equation}
\end{problemstate}

In this problem statement, the retrained posterior $p(Z|X_r)$ can be obtained by retraining the IB model based on the remaining dataset as 
\begin{equation}
	\begin{aligned}
		\min \  & I(X_r;Z), \\
		\text{s.t. } \ & I(X_r;Y_r) - I(Z;Y_r) = I(X_r;Y_r|Z)=0.
	\end{aligned}
\end{equation}
\Cref{eq_eu} means that an unlearned IB-trained model is still hoped to keep the property like a retrained model, squeezing as much information of $X_r$ as possible from the representation $Z$ while keeping a good utility of approximating $Y_r$. Retraining the model from scratch remains a viable approach under this definition of unlearning. However, as previously discussed, retraining from scratch incurs significant storage and computational costs and has no protection to defend against reconstruction attacks on unlearning. Therefore, we propose the CRFU approach to solve this problem in the following.

\begin{figure}
	\centering
	\includegraphics[width=0.98\linewidth]{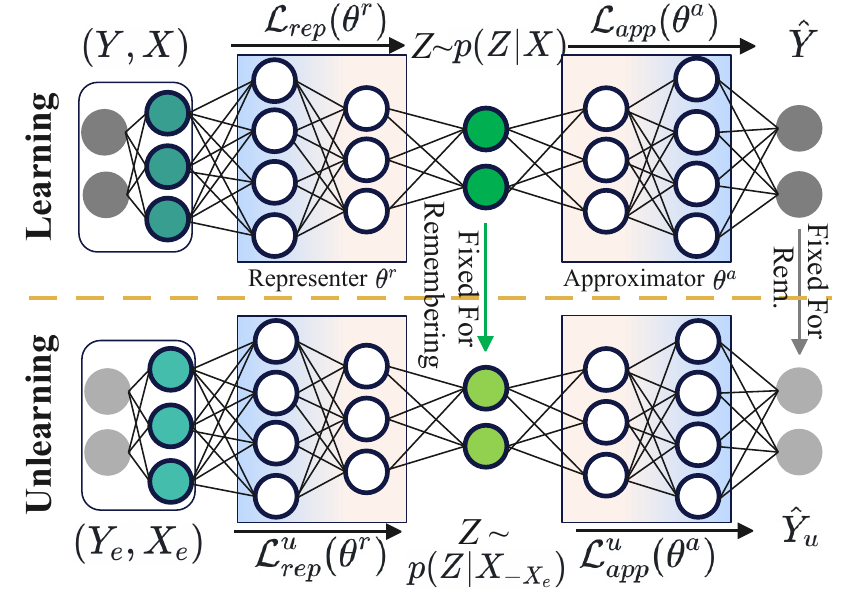}
	\vspace{-2mm}
	\caption{IB learning process (upper half) and CRFU unlearning process (lower half) with a fixed learned IB model (a trained and fixed upper half).}
	\label{fig:unlvibrfu}
\end{figure}

\subsection{CRFU Method} \label{detailed_rfu}

\subsubsection{Overview of CRFU}
We begin by briefly introducing the core concept of the proposed CRFU, which is illustrated in \Cref{fig:unlvibrfu}. CRFU is tailored to unlearn models that were trained using the IB method (shown in the upper half in \Cref{fig:unlvibrfu}). An IB method mainly learns a representation $Z$ that compresses information of inputs while retaining sufficient information for the targets. The CRFU, the lower half in \Cref{fig:unlvibrfu}, aims to erase the influence of $D_e=(X_e, Y_e)$ from $Z$. To achieve this goal, we minimize the mutual information $I(X_e;Z)$ and $I(Y_e;Z)$; however, directly minimizing them could result in the representation $Z$ catastrophically forgetting everything, decreasing model utility. To mitigate the degradation of model utility, we introduce a constraint ensuring that the unlearned representation $Z$, derived from the distribution $p(Z|X_{-X_e})$, should be close to the representation $Z$ obtained from the original-fixed distribution $p(Z|X)$. We deal with the predicting approximation $\hat{Y}$ in the same way. Finally, combining the minimizing mutual information and a constraint, we formalize the losses $\mathcal{L}^u_{rep}$ and $\mathcal{L}^u_{app}$ to unlearn the trained IB model.

\subsubsection{Theoretical Exact Compressive Representation Forgetting Unlearning}

In an IB model, the objective is to learn a representation $Z$ by minimizing the mutual information $I(X;Z)$, while simultaneously maximizing $I(Y;Z)$. In our CRFU approach, we specifically focus on minimizing $I(X_e;Z)$ for the inputs $X_e$ designated for erasure. This is aimed at effectively eliminating the information of $X_e$ from the previously learned representation $Z$. Similarly, we work on minimizing $I(Y_e;Z)$, the mutual information between the targets $Y_e$ and the representation $Z$, thereby removing the information of $Y_e$ from $Z$. This dual minimization ensures a comprehensive unlearning of both input and label information from the representation.

The ideal unlearned representation should contain no information about $(X_e, Y_e)$, optimally satisfying the conditions where $I(X_e;Z)=0$ and $I(Y_e;Z)=0$, indicating zero mutual information between the erased data and the representation $Z$. However, directly minimizing $I(X_e;Z)$ and $I(Y_e;Z)$ poses a significant risk: it could lead the representation $Z$ to inadvertently ``forget'' everything, including the valuable knowledge it had previously learned. Therefore, we must keep the unlearned representation $Z$ and approximation $\hat{Y}$ remembering the knowledge learned in the former training.

To maintain the learned knowledge, we propose a solution that minimizes the Kullback–Leibler divergence (KLD) \cite{joyce2011kullback} between the unlearned and originally learned representations. This involves minimizing both $\text{KL}[p(Z|X) ||p(Z|X_{-X_e})]$ and $\text{KL}[p(\hat{Y}|Z)||p(\hat{Y}_{-(X_e,Y_e)}|Z)]$, ensuring that the unlearned representations remain closely aligned with the original ones. During unlearning, $p(Z|X)$ and $p(\hat{Y}|Z)$ are calculated using the trained IB model as a temporary reference model. Combining the above solutions, we formalize the unlearned representation loss as 
\begin{equation} \label{rep_unl}
	\mathcal{L}_{rep}^u =  I(X_e;Z) + \text{KL}[p(Z|X) ||p(Z|X_{-X_e})],
\end{equation}
and the unlearned approximation loss as
\begin{equation} \label{app_unl}
	\mathcal{L}_{app}^u =I(Y_e;Z) +\text{KL}[p(\hat{Y}|Z)||p(\hat{Y}_{-(X_e,Y_e)}|Z)].
\end{equation}
Integrating these two loss functions, we formulate the optimization of CRFU as the proposition outlined below.

\begin{proposition} \label{unl_eq_pro} 
	Define the CRFU loss function as
	\begin{equation}\label{rfu_loss} 
		\begin{aligned}
				\mathcal{L}^u &= \beta \cdot \mathcal{L}_{rep}^u + \mathcal{L}_{app}^u  \\
				&= \beta \cdot (  I(X_e;Z) + \text{KL}[p(Z|X) ||p(Z|X_{-X_e})] )  \\
				&+ I(Y_e;Z) + \text{KL}[p(\hat{Y}|Z)||p(\hat{Y}_{-(X_e,Y_e)}|Z)] .
		\end{aligned}
	\end{equation}
Then, minimizing the loss in \Cref{rfu_loss} to unlearn the erased dataset $D_e$ from an IB model trained based on $D$ is equivalent to retraining an IB model by minimizing $\mathcal{L}_r = \beta I(X_r;Z) - I(Y_r;Z)$ based on $D_r$.  
\end{proposition}


Since CRFU unlearns an IB model trained through \Cref{eq:total_loss_q}, we can combine the original IB loss and CRFU loss to prove that minimizing \Cref{rfu_loss} based on a trained IB model is equivalent to minimizing the loss of retraining an IB model on the remaining dataset. 
We first prove the representation loss. Given that $X = X_r \cup X_e $ and $X_r \cap X_e = \emptyset$, and considering that the data are IID, along with the conditional independence of $X_r$ and $X_e$ given $Z$, it follows that $I(X;Z) = I(X_r;Z) + I(X_e;Z)$. Since $ I(X_e;Z)\geq 0$ and $\text{KL}[p(Z|X) ||p(Z|X_{-X_e})]\geq 0$, we can expand original and unlearned representation loss $\mathcal{L}_{rep} + \mathcal{L}_{rep}^u$ function as 
\begin{equation} \label{rep_u_proof}
	\small
	\begin{aligned}
		\mathcal{L}_{rep} + \mathcal{L}_{rep}^u&=  I(X;Z) +  I(X_e;Z)+  \text{KL}[p(Z|X) ||p(Z|X_{-X_e})] \\
		&= I(X_r;Z) + 2I(X_e;Z) + \text{KL}[p(Z|X) ||p(Z|X{-X_e})]  \\
		&\geq I(X_r;Z) .
	\end{aligned}
\end{equation}
It is clear that the sum of the representation loss $\mathcal{L}_{rep}$ and the unlearned representation loss $\mathcal{L}_{rep}^u$ serves as an upper bound for the mutual information $I(X_r;Z)$. Minimizing this upper bound during both original training and unlearning is equivalent to minimizing $I(X_r;Z)$ itself.

Similarly, for the original and unlearned approximation loss function $\mathcal{L}_{app} + \mathcal{L}_{app}^u$, we can obtain the expanded approximation as
\begin{equation} \label{app_u_proof}
	\begin{aligned}
		\mathcal{L}_{app} + \mathcal{L}_{app}^u &=-I(Y;Z) +  I(Y_e;Z)\\
		&+  \text{KL}[p(\hat{Y}|Z)||p(\hat{Y}_{-(X_e,Y_e)}|Z)] \\
		&= -(I(Y;Z) - I(Y_e;Z)) \\
		&+ \text{KL}[p(\hat{Y}|Z)||p(\hat{Y}_{-(X_e,Y_e)}|Z)] \\
		&= -I(Y_r;Z) + \text{KL}[p(\hat{Y}|Z)||p(\hat{Y}_{-(X_e,Y_e)}|Z)] \\
		& \geq -I(Y_r;Z).
	\end{aligned}
\end{equation}
Since $Y=Y_r \cup Y_e$ and $Y_r \cap Y_e = \emptyset$ and the data are IID, it follows that $I(Y_r;Y_e)=0$ and $H(Y) = H(Y_r,Y_e) = H(Y_r) + H(Y_e) - I(Y_r;Y_e)=H(Y_r) + H(Y_e)$. We can similarly achieve $H(Y_r|Z)$. And since $I(Y_r;Z) = H(Y_r) - H(Y_r|Z)$, thus \Cref{app_u_proof} holds.
Minimizing the original and unlearned approximation as \Cref{app_u_proof} is the upper bound of $-I(Y_r;Z)$ in retraining.
Therefore, minimizing the CRFU loss function based on a trained IB model is equivalent to minimizing the loss function of retraining an IB model based on the remaining dataset.

\subsubsection{Variational Optimization Method for CRFU}
Optimizing the CRFU loss function \Cref{rfu_loss} in a deep learning scenario with a huge amount of training data is challenging. To address this issue, we present a variational method to make the CRFU loss function calculatable in deep learning. The unlearned representation loss can be expanded as 
\begin{equation} \label{unl_rep_p}
	\small
	\begin{aligned}
		\mathcal{L}_{rep}^u&=  I(X_e;Z) + \text{KL}[p(Z|X) ||p(Z|X_{-X_e})] \\
		& =  \text{KL}[p_{\theta^r}(Z|X_e)|| p_{\theta^r}(Z)] +   \text{KL}[p_{\theta_{fix}^r}(Z|X)|| p_{\theta^r}(Z|X_{-X_e})]     .
	\end{aligned}
\end{equation} 
Generally, minimizing this loss is intractable due to the complexity involved in computing the KL term. This computation requires knowledge of the marginal distribution $Z$ and $p_{\theta^r}(Z) = \int dx \ p(z|x)p(x)$, which is not easily obtainable. To address this issue, we introduce $q(Z)$ as a variational approximation to this marginal as the variational IB learning in \cite{alemi2016deep, achille2018information}. Since $\text{KL}[p(Z)||q(Z)] \geq 0 \Longrightarrow \int dz p(z) \log p(z) \geq \int dz p(z) \log q(z)$, we have the following upper bound of \Cref{unl_rep_p}: 
\begin{equation} \label{unl_rep_p_q}
	\begin{aligned}
		\mathcal{L}_{rep}^u&=  I(X_e;Z) + \text{KL}[p(Z|X) ||p(Z|X_{-X_e})] \\
		& \leq \underbrace{  \text{KL}[p_{\theta^r}(Z|X_e)|| q(Z)]}_{\text{Forgetting the erased inputs from the representation}} \\
		& + \underbrace{  \text{KL}[p_{\theta_{fix}^r}(Z|X)|| p_{\theta^r}(Z|X_{-X_e})] }_{\text{Remembering the original learned representation}}.
	\end{aligned}
\end{equation}
This unlearned representation loss can be explained as forgetting the impact of the erased data's inputs (first term) and remembering the knowledge of original full training inputs (second term). The proof of the correctness of \Cref{unl_rep_p_q} is similar to the proof of Eq. (14) in \cite{alemi2016deep}. Hence, we omit the detailed proof here.

Following the solutions to $\mathcal{L}_{app}$ in \cite{alemi2016deep, achille2018information}, the unlearned approximation loss function can be optimized in a similar way as
\begin{equation}\label{app_expand}
	\begin{aligned}
		\mathcal{L}_{app}^u&=  I(Y_e;Z) + \text{KL}[p(\hat{Y}|Z)||p(\hat{Y}_{-(X_e,Y_e)}|Z)] \\
		&\simeq \underbrace{  \int dz \ p(Z|X_{-X_e}) \log \  p_{\theta^a}(Y_e|Z) }_{\text{Forgetting the erased targets from the approximation}} \\
		& + 	\underbrace{  \text{KL}[p_{\theta_{fix}^a}(\hat{Y}|Z)||p_{\theta^a}(\hat{Y}_{-(X_e,Y_e)}|Z)] }_{\text{Remembering the original learned approximation} }
	\end{aligned}
\end{equation}
The computable optimization loss \Cref{app_expand} represents a balance between completely unlearning the information of the erased labels $Y_e$ (first term) and not entirely forgetting the originally learned information of $Y$ in representation $Z$ (second term). 
Combining \Cref{app_expand,unl_rep_p_q} and optimizing them together is the CRFU loss \Cref{rfu_loss} that we proposed in proposition \ref{unl_eq_pro}. When minimizing this loss, we should notice that the second term in \Cref{app_expand,unl_rep_p_q} is calculated based on the original full training dataset; however, storing all original datasets and training it again is impractical. A simple way we used is fixing the trained model (i.e., the model before unlearning) as a temp model that is specially used to calculate the first term but only based on the erased dataset.

\subsubsection{An Unlearning Rate for CRFU}
To make the unlearning method adaptive to different tasks, we introduced an unlearning rate parameter, $\beta_u$, which serves to regulate the extent of unlearning during both the representation and approximation unlearning process. 
Specifically, we add an unlearning rate parameter $\beta_u$ before the forgetting terms in \Cref{unl_rep_p_q,app_expand} to optimize the balance between forgetting the information of erased samples and remembering the representation learned before.
Adjusting the unlearning rate $\beta_u$ will also impact the unlearning speed. 
The final equation can be described as 
\begin{equation}\label{final_vibu}
	\small
	\begin{aligned}
		\mathcal{L}^u & = \beta \cdot \mathcal{L}_{rep}^u(\beta_u) +  \mathcal{L}_{app}^u(\beta_u) \\
		& \simeq  \beta \cdot (  \beta_u \cdot  \text{KL}[p_{\theta^r}(Z|X_e)|| q(Z)]  \\
		& + \text{KL}[p_{\theta^r}(Z|X)|| p_{\theta_{fix}^r}(Z|X_{-X_e})] )\\
		& + \beta_u \cdot \int dz \ p(Z|X_{-X_e}) \log \  p_{\theta^a}(Y_e|Z)  \\
		& +  \text{KL}[p_{\theta_{fix}^a}(\hat{Y}|Z)||p_{\theta^a}(\hat{Y}_{-(X_e,Y_e)}|Z)]
	\end{aligned}
\end{equation}

We present a pseudocode of CRFU in Algorithm \ref{alg_RFUnL}.
At the beginning of executing representation forgetting unlearning, we first prepare a trained IB model $\mathcal{M}(\theta^r,\theta^a)$, the erased dataset $D_e=(X_e, Y_e)$, and training epochs $E$. Then, we fix the original optimal model $\mathcal{M}_{fix}(\theta^r_{fix}, \theta^a_{fix}) \gets \mathcal{M}(\theta^r,\theta^a)$ as a temp model for the later unlearned representation and approximation loss calculation, as shown in Line 1 of Algorithm \ref{alg_RFUnL}.
When executing unlearning training, lines 2 to 7, we draw a minibatch of $m$ data samples $\{(x_i, y_i)\}_{i=1}^m$ from $D_e$. Using the representer $\theta^r$, we generate the corresponding representations $z_i$ for these samples. The unlearning loss function is then calculated as per \Cref{unl_rep_p_q,app_expand}. Following this, we update the parameters of the representer and approximator $(\theta^r, \theta^a)$ in the IB model in accordance with \Cref{final_vibu} at line 7.

\begin{algorithm}[t]
	\caption{Compressive Representation Forgetting Unlearning} \label{alg_RFUnL}
	\begin{normalsize}
		\BlankLine
		\KwIn{The trained model $\mathcal{M}(\theta^r, \theta^a)$, the erased dataset $(X_e, Y_e)$ and training epochs $E$}
		\KwOut{Unlearned model $\mathcal{M}_{u}(\theta^r_u,\theta^a_u)$}
		Establish the fixed temporary model: $\mathcal{M}_{fix}(\theta^r_{fix}, \theta^a_{fix}) \gets \mathcal{M}(\theta^r,\theta^a)$\\
		\For{$E$ epochs}{
			Draw a minibatch of $m$ samples $\{(x^i, y^i)\}_{i=1}^m$ from erased dataset $D_e=(X_e,Y_e)$; \\
			Generate $z^i \backsim p_{\theta^r}(\cdot|x^i)$;\\
			Compute the loss function \Cref{unl_rep_p_q} for unlearning representation on a per-sample basis 
			$\mathcal{L}_{rep}^u(\beta_u) =  \frac{\beta_u}{m} \sum_{i=1}^{m}  \text{KL}  [p_{\theta^r}(Z|x^i) ||  \prod_{j}^{|Z|} q^i_j(z^i_j)] $ \newline \hspace*{3em} $ +  \frac{1}{m} \sum_{i=1}^m  \text{KL}  [p_{\theta_{fix}^r}(Z|x^i) || p_{\theta^r}(Z|x^i) ]  $ \\
			Compute the loss function \Cref{app_expand} for unlearning approximation on a per-sample basis
			$\mathcal{L}_{app}^u(\beta_u) =   \beta_u \cdot \frac{1}{m} \sum_{i=1}^{m} \log\ p_{\theta^a}(y^i| z^i) $ \newline \hspace*{3em} $ +  \frac{1}{m} \sum_{i=1}^m \text{KL}[p_{\theta_{fix}^a}(\hat{y^i}|z^i)||p_{\theta^a}(\hat{y^i}_{-(X_e,Y_e)}|z^i)] $ \\
			Update the model based on the gradients of the integrated loss functions \Cref{final_vibu} as
			$ (\theta^r, \theta^a) \gets (\theta^r, \theta^a) - \eta \nabla _{(\theta^r, \theta^a)} (\beta \cdot \mathcal{L}_{rep}^u(\beta_u) $ \newline \hspace*{3em} $+  \mathcal{L}_{app}^u(\beta_u)  ) $  \\
		}
		Return $\mathcal{M}_{u}(\theta^r,\theta^a)$;
	\end{normalsize}
\end{algorithm}

\section{Theoretical Analysis of Privacy Leakage Defense of CRFU}
\label{theoretical_a}

In this section, we give a theoretical explanation of why our proposed CRFU can effectively defend against privacy leakage attacks of unlearning in a black-box ML setting. 

\subsection{Reasons behind Effective Privacy Leakage Attacks}
Before discussing why our proposed CRFU is effective against privacy leakage attacks, we first explain why the existing privacy inference attacks can effectively recover users' privacy. Similar to most model inference attacks, the privacy inference attack on unlearning aims to recover training samples by analyzing a black-box ML model's outputs. These attacks focus on differences between the outputs of the original and updated unlearning models to reconstruct the privacy of the deleted data in this unlearning process, as discussed in \cite{chen2021machine, gao2022deletion,Hu2024sp}. The effectiveness of such attacks lies in the requirement that the unlearning mechanism must ensure the unlearned model forgets the specified samples. Therefore, effective unlearning, which erased the information of specified data from the model, will result in different outputs for the same inputs before and after unlearning. Attackers can exploit private information from erased data by comparing the model's different outputs before and after unlearning. They mimic the changes between the original and unlearned models using an IID auxiliary dataset, thereby training a model to infer the privacy of the erased data.

\subsection{How CRFU Defends against Privacy Leakage Attacks}
CRFU can effectively defend against such privacy leakage attacks because both learning and unlearning are based on the IB framework, which discards maximized information of inputs $X$ of the bottleneck $Z$, leaving little information in the model output for adversaries to infer. Specifically, CRFU implements the data erasure of inputs $X_e$ by further minimizing $I(X_e;Z)$ based on the minimized $I(X;Z)$. Ideally, we denote the information remaining after unlearning as $I(X_r;Z)$. Therefore, the maximum information that an adversary can infer from the different representations $Z$ before and after unlearning is $I(X;Z) - I(X_r;Z) = I(X_e;Z)$. During the learning process, the $I(X_e;Z)$ is minimized with a constraint of maintaining enough information for targets, i.e., $I(X_e;Z) \geq I(Y_e;Z)$. Therefore, the ideal protection of CRFU is $I(X_e;Z) = I(Y_e;Z)$. Attackers are only able to infer information about the labels, but they cannot reconstruct the samples that have been erased.

It can also be explained from the perspective of a Markov chain, which is applied to both the model learning and unlearning processes. In the IB model training, the Markov chain is established as $Y \to X \to Z \to \hat{Y}$. For the CRFU training, the corresponding chain is $Y_e \to X_e \to Z \to \hat{Y}_{-(X_e,Y_e)}$. 
It ensures that attackers cannot infer more information from $\hat{Y}$ than $Z$ in a black-box ML scenario because $\hat{Y}$ is derived from $Z$. Due to the Markov chain principle, once information is lost in one layer, it cannot be regained in subsequent layers. This process can be described as
\begin{equation}\label{mutual information}
	\begin{cases}
		I(X;Y) \geq I(Z;Y) \geq I(\hat{Y}, Y) \\
		I(X_e;Y_e) \geq I(Z;Y_e)  \geq I(\hat{Y}_{-(X_e,Y_e)}, Y_e)
	\end{cases}
\end{equation}
Therefore, for an IB model that has undergone unlearning through CRFU, the upper bound of privacy inference attacks based on the different outputs is the reconstruction capability on the different representation $Z$. This is because the outputs of the original trained model ($\hat{Y}$) and the unlearned model ($\hat{Y}_{-(X_e,Y_e)}$) are derived from the representations before and after unlearning, respectively. The information that the adversary can infer will be no more than $I(Y_e;Z)$.

\subsection{$\beta$-Compression Defense of CRFU}
We provide an example of the representation term's process using the Gaussian distribution case, which is widely employed in many studies \cite{kusner2017grammar,alemi2016deep}. Let the assumed prior $q(Z) = \mathcal{N}(Z;0, \textbf{I})$ and the posterior $p(Z|X) = \mathcal{N}(Z; \mu^i, \sigma^i)$ are Gaussian. Let $J$ be the dimensionality of $Z$, $\mu^i$ and $\sigma^i$ are the mean and standard deviation output by the representer $\theta^r$ at datapoint $x^i$, and let $\mu^i_j$ and $\sigma^i_j$ denote the $j$-th element of these vectors. Then we have  
\begin{equation} \label{Gaussian_kl}
	\text{KL}[p_{\theta^r}(Z|x^i) || q_{\theta^r}(Z)] =  \frac{1}{2} \sum_{j=1}^{J} ((\mu^i_j)^2 + (\sigma^i_j)^2 - \log ((\sigma^i_j)^2 ) - 1  )
\end{equation}
The KL divergence of \Cref{Gaussian_kl}  is one optimizing term of \Cref{eq:total_loss_q,final_vibu}. And the mean $\mu^i$ and s.d. $\sigma^i$ are determined by $x^i$ and the representer parameters $\theta^r$. The representer network $\theta^r$ transforms the input data $x^i$ into the parameters of a complex, high-dimensional joint Gaussian distribution. Specifically, for each input $x^i$, the representer outputs a mean vector $\mu^i_{J} = \{\mu^i_{1}, \mu^i_{2}, ..., \mu^i_{j} \} $ and a log-variance vector $\log( (\sigma^{i}_{J})^2 ) = \{\log( (\sigma^{i}_{1})^2 ), \log( (\sigma^{i}_{2})^2 ),..., \log( (\sigma^{i}_{j})^2 )\} $. Here, for each input, the representer $\theta^r$ will output $J$ joint Gaussian distribution. These parameters describe a high-dimensional Gaussian distribution for the data point $x^i$ as 
\begin{equation}
	\mu^i_J, \log ( (\sigma^i_J)^2) = \theta^r (x^i).
\end{equation}
Then, we can calculate the representation $Z^i_J$ using the reparameterization trick. We sample a latent vector $Z_{j}^i$ from this distribution:
\begin{equation}
	Z^i_J = \mu^i_J + \sigma^i_J \cdot \epsilon,
\end{equation}
where $\epsilon \sim \mathcal{N} (0, I)$. When \Cref{Gaussian_kl} is minimized, such as in the case of $\text{KL}[p_{\theta^r}(Z|x^i) || q_{\theta^r}(Z)] = 0$, the learned representation posterior $p(Z|x^i)$ becomes identical to the assumed prior, $\mathcal{N}(Z; 0, \textbf{I})$.  In this scenario, the representation $Z$ loses all information about the sample $x^i \in X$ and $x^i \in X_e$, limiting attackers to inferring only the information of the general prior, $\mathcal{N}(Z; 0, \textbf{I})$. Greater distortion thus offers a stronger defense against privacy inference attacks.

However, to ensure model utility, there is a constraint that $Z$ must contain sufficient information about $Y$. A distortion rate $\beta$ is introduced in \Cref{eq:total_loss_q,final_vibu} to balance the information compression and the model utility. We can define the protection capability of CRFU as follows:
\begin{defn}
CRFU achieves the \textbf{$\beta$-Compression Defense} against the privacy leakage attacks on unlearning if the original IB model is trained using \Cref{eq:total_loss_q} and CRFU unlearns the trained IB model using \Cref{final_vibu}, where \Cref{eq:total_loss_q,final_vibu} are optimized using the same compressive parameter $\beta$.
\end{defn}
Since the compression ratio is controlled by $\beta$ in \Cref{eq:total_loss_q,final_vibu}, an increasing $\beta$ leads to greater information distortion of $X$ and a reduced privacy inference effect from $Z$ while concurrently diminishing the model utility about $Y$. To balance the tradeoff between prediction utility and defense against privacy leakage attacks, selecting a suitable $\beta$ is crucial. Our upcoming experiments will further assess the defense effectiveness against privacy leakage attacks at varying levels of $\beta$.

\begin{table*}[h]
	\centering
	\caption{Reconstructing quality of different $\beta$ \vspace{-2mm}}
	\label{table_reconstruct}
	\begin{tabular}{ |l|c|c|c|c|c|c| }
		\hline
		Dataset & Original Images & \makecell[c]{$\beta=0$ (Normal\\unl. models\\HBU and VBU)}& $\beta=0.001$ & $\beta=0.01$ & $\beta=0.1$ & $\beta=1$ \\ 
		\hline
		\makecell[l]{MNIST}
		&
		\begin{minipage}[t]{0.1\textwidth}
			\centering    
			\includegraphics[width=2cm]{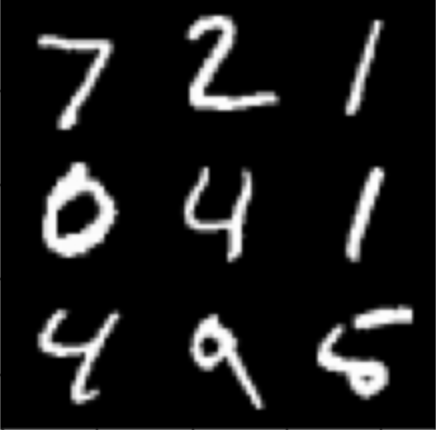}  
		\end{minipage}     &
		$\includegraphics[width=2cm]{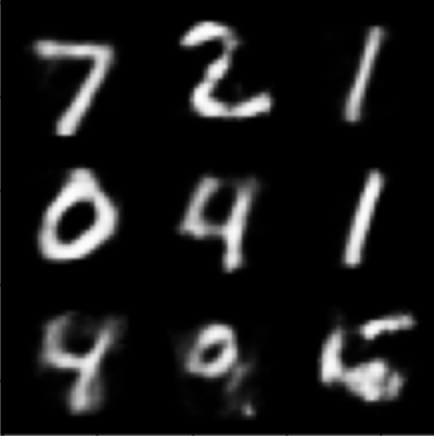}$
		&
		$\includegraphics[width=2cm]{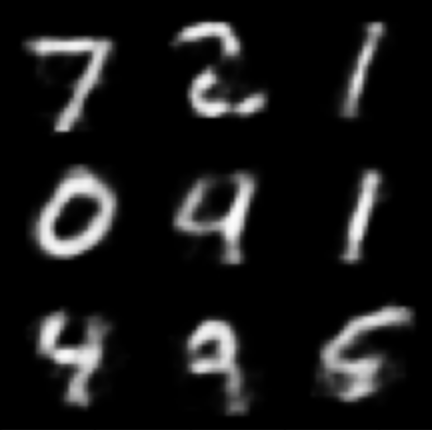}$ 
		&
		$\includegraphics[width=2cm]{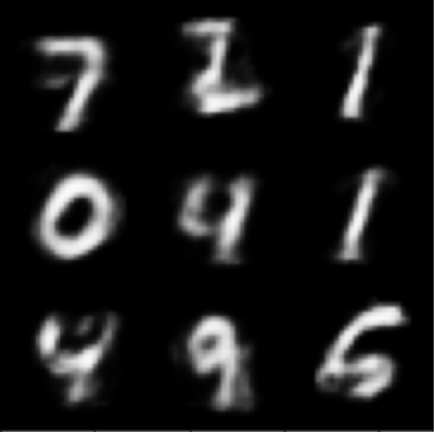}$  
		&
		$\includegraphics[width=2cm]{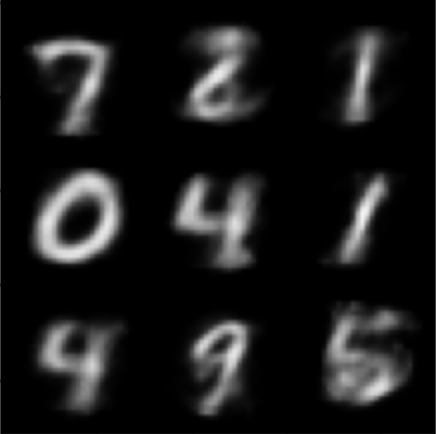}$  
		&
		$\includegraphics[width=2cm]{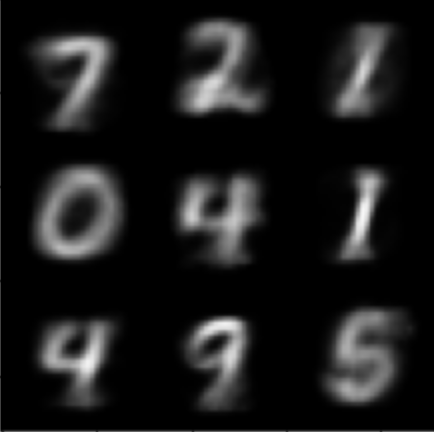}$  
		\\ 
		\hline
		Reconstruction MSE & - & 231.31 & 252.23 & 312.87 & 478.76& 668.07 \\  		
		\hline
		Inference AUC &- & 0.674 & 0.653 &0.573 & 0.526 & 0.515 \\
		\hline
		KLD to Prior $q(Z)$ & - & 12.39 & 5.00 & 1.45 & 0.37& 0.14 \\ 
		\hline
		Accuracy & - & $97.45\%$ & $97.32\%$ & $96.99\%$  & $96.45\%$ & $95.75\%$ \\ 
		\hline
		Fashion MNIST&
		\begin{minipage}[t]{0.1\textwidth}
			\centering    
			\includegraphics[width=2cm]{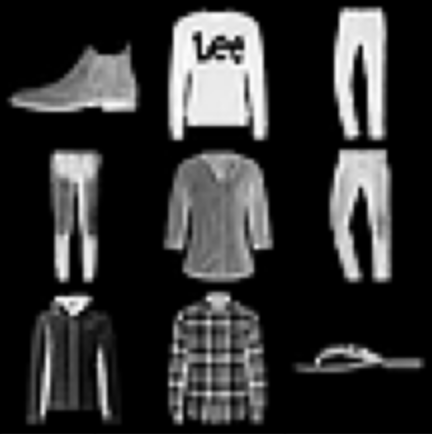}  
		\end{minipage}     &
		\begin{minipage}[t]{0.1\textwidth}
			\centering    
			\includegraphics[width=2cm]{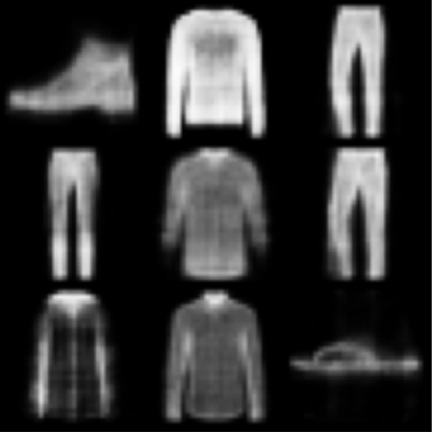}  
		\end{minipage} &
		\begin{minipage}[t]{0.1\textwidth}
			\centering    
			\includegraphics[width=2cm]{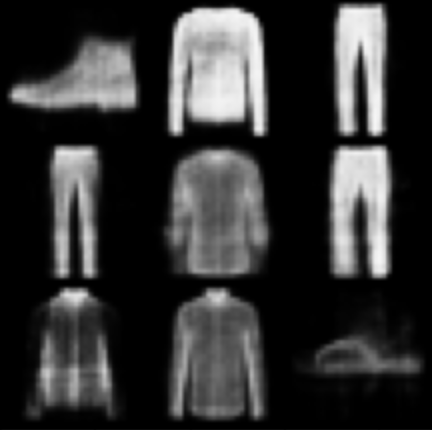}  
		\end{minipage}   &
		\begin{minipage}[t]{0.1\textwidth}
			\centering    
			\includegraphics[width=2cm]{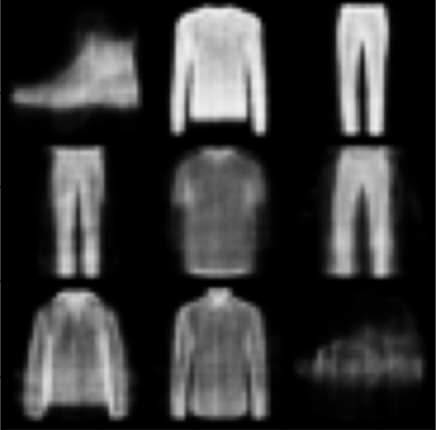}  
		\end{minipage} &
		\begin{minipage}[t]{0.1\textwidth}
			\centering    
			\includegraphics[width=2cm]{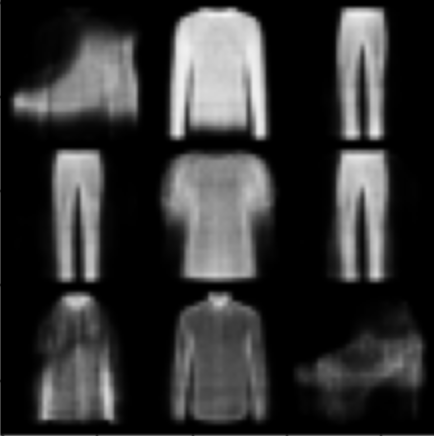}  
		\end{minipage} &
		\begin{minipage}[t]{0.1\textwidth}
			\centering    
			\includegraphics[width=2cm]{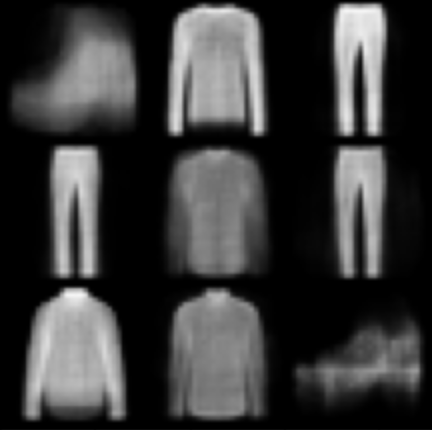}  
		\end{minipage}\\ 
		\hline
		Reconstruction MSE & - & 220.31 & 240.41 & 284.79 & 386.08& 561.73 \\  		
		\hline
		Inference AUC &- & 0.656 & 0.638 &0.579 & 0.519 & 0.515 \\
		\hline
		KLD to Prior $q(Z)$ & - & 12.23 & 5.77 & 1.78 & 0.33& 0.13 \\ 
		\hline
		Accuracy & - & $87.95\%$ & $87.72\%$ & $87.29\%$  & $87.15\%$ & $86.15\%$ \\ 
		\hline
		CIFAR10&
		\begin{minipage}[t]{0.1\textwidth}
			\centering    
			\includegraphics[width=2cm]{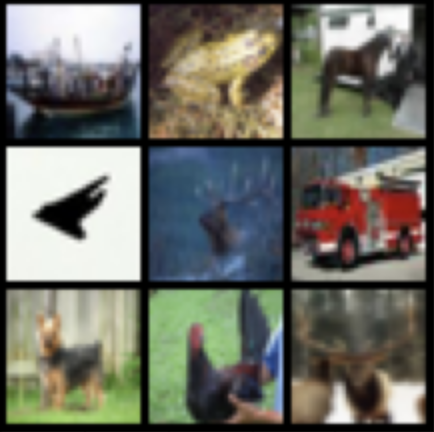}  
		\end{minipage}     &
		\begin{minipage}[t]{0.1\textwidth}
			\centering    
			\includegraphics[width=2cm]{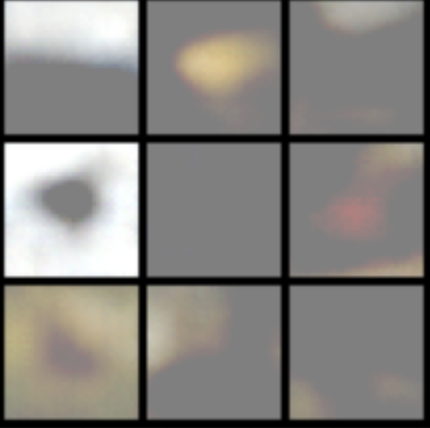}  
		\end{minipage} &
		\begin{minipage}[t]{0.1\textwidth}
			\centering    
			\includegraphics[width=2cm]{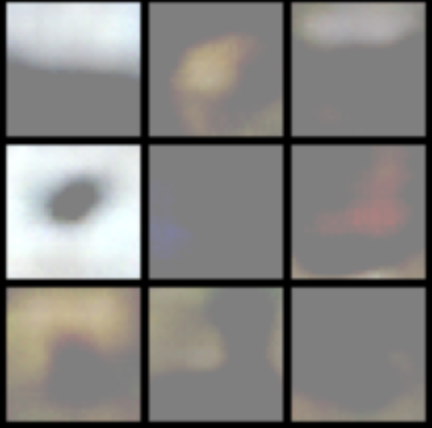}  
		\end{minipage}   &
		\begin{minipage}[t]{0.1\textwidth}
			\centering    
			\includegraphics[width=2cm]{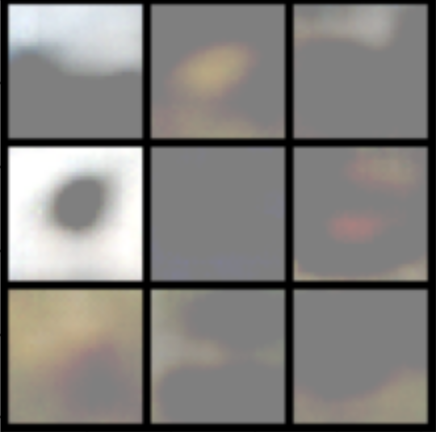}  
		\end{minipage} &
		\begin{minipage}[t]{0.1\textwidth}
			\centering    
			\includegraphics[width=2cm]{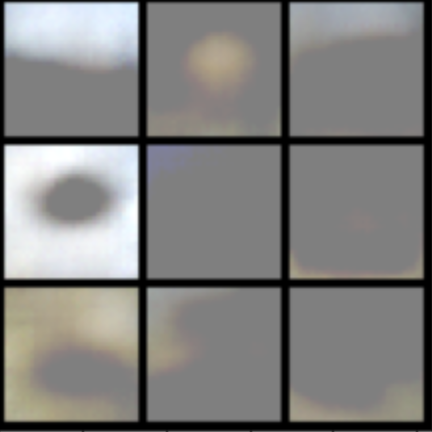}  
		\end{minipage} &
		\begin{minipage}[t]{0.1\textwidth}
			\centering    
			\includegraphics[width=2cm]{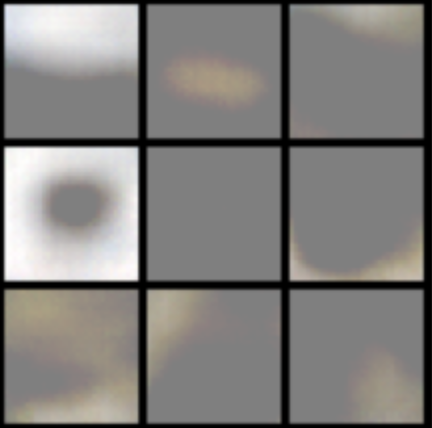}  
		\end{minipage}\\ 
		\hline
		Reconstruction MSE & - & 1174.33 & 1180.84 & 1188.57 & 1267.78& 1550.24 \\  		
		\hline
		Inference AUC &- & 0.821 & 0.810 &0.769 & 0.669 & 0.571 \\
		\hline
		KLD to Prior $q(Z)$& - & 9.09 & 3.75 & 1.76 & 0.53& 0.11 \\ 
		\hline
		Accuracy & - & $84.75\%$ & $84.75\%$ & $84.00\%$  & $83.96\%$ & $82.95\%$ \\ 
		\hline
	\end{tabular}
\end{table*}

\section{Performance Evaluation}
\label{ex}
\subsection{Experiment Setup}

\noindent 
\textbf{Datasets.}
We assess the effectiveness of the proposed Compressive Representation Forgetting Unlearning (CRFU) method using four benchmark datasets: MNIST, Fashion-MNIST \cite{xiao2017fashion}, CIFAR10 \cite{krizhevsky2009learning}, and STL-10 \cite{coates2011analysis}. The four datasets are benchmark datasets for image classification tasks, which cover a wide range of object categories with different learning complexities.

\noindent
\textbf{Models.} In our experiments, we utilize two model architectures of different sizes, a 5-layer multi-layer perceptron (MLP) with ReLU activations and ResNet-18, to construct the IB model. Specifically, we employ two 5-layer MLP models, one as the representer and one as the approximator, to construct the IB models trained on MNIST and Fashion-MNIST. We employ one ResNet-18 as the representer and one 5-layer multi-layer MLP as the approximator to construct the IB models trained on CIFAR10 and STL-10.

For experimental simplicity, we set a consistent minibatch size of $m=20$. We set the learning rate $\eta = 0.001$ on both the MNIST and Fashion MNIST datasets. For experiments conducted on the CIFAR10 and STL-10 datasets, we employ the learning rate to $\eta=0.0005$. In the performance evaluation for defense, we set the unlearning rate $\beta_u=0.1$. In the evaluation of unlearning utility, we set the distortion rate $\beta=0.001$ on MNIST and Fashion MNIST and $\beta=0.0001$ on CIFAR10 and STL-10. All methods were implemented using PyTorch and tested on a computing cluster equipped with four NVIDIA 1080ti GPUs.

\noindent
\textbf{Evaluation Metrics for Attacking Methods and Unlearning Benchmarks.}
Our evaluation focuses on two aspects: the defense against privacy leakage attacks and the utility of the unlearned model. 
From the perspective of defense capabilities, we test all unlearning methods against the state-of-the-art attack implementations as described in \cite{salem2020updates,chen2021machine}, including the reconstruction attack and the membership inference attack. For an effective attack, we implement the reconstruction and membership inference attacks directly on the representation $Z$, representing the upper bound of such attacks in CRFU, based on the outputs $\hat{Y}_{-(X_e,Y_e)}$. As discussed in Section \ref{theoretical_a}, an attacker cannot infer more information from $\hat{Y}$ and $\hat{Y}_{-(X_e,Y_e)}$ than from $Z$ before and after unlearning, since both the original model and CRFU base their outputs on $Z$. Following the approach in \cite{salem2020updates}, we measure the quality of reconstruction using mean square error (MSE). We rely on the traditional AUC metric to measure the absolute performance of the membership inference according to \cite{chen2021machine}.

From the perspective of unlearned model utility, we compare the effectiveness and efficiency of CRFU and state-of-the-art two main kinds of approximate unlearning methods, including Hessian-matrix-based unlearning (HBU) \cite{sekhari2021remember,guo2019certified} and variational bayesian unlearning (VBU) \cite{fu2022knowledge, nguyen2020variational}. To rigorously assess the effectiveness of unlearning methods, we adopt a widely-used technique as outlined by \citeauthor{hu2022membership}\cite{hu2022membership}, which involves embedding backdoor triggers into the samples that are to be erased during the initial training of the original IB model. The objective of all unlearning methods is to eliminate the influence of these embedded backdoors from the trained models. After unlearning, we evaluate the success of these methods by checking whether the backdoor still poses a threat to the unlearned model. The effectiveness of the unlearning methods is gauged in two ways: firstly, by measuring the model's accuracy on a test dataset, and secondly, by assessing the backdoor accuracy on the erased dataset \cite{hu2022membership}. Additionally, we analyze the efficiency of unlearning by timing the model's run, computed as the product of the per-batch training time and the total number of training epochs.

\subsection{Evaluations of Defense Capability}

As analyzed in Section \ref{theoretical_a}, CRFU performs unlearning based on a trained IB model. 
Increasing the value of $\beta$ during training leads to a learned representation that more closely approximates the general prior. A smaller KLD, $\text{KL}[p(Z|X)||q(Z)]$, implies better distortion quality. However, increased distortion can compromise prediction accuracy. To explore the relationship between $\beta$ and its defense effectiveness, we conducted experiments with varying $\beta$ values in both the original model and the CRFU on the MNIST, Fashion MNIST, and CIFAR10 datasets. For clarity of illustration, we have not added backdoors in the erased data during this experiment. We set a fixed unlearning rate $\beta_u=0.1$, and only 9 erased samples of the full training data. To facilitate the experimental process easily, we carried out the reconstruction and membership inference attack on the representation $Z$, marking the highest level of attack an adversary can achieve with $\hat{Y}$ and $\hat{Y}_{-(X_e,Y_e)}$. The results showcasing the reconstruction effect and model utility across different $\beta$ values are presented in Table \ref{table_reconstruct}.

In our study on the MNIST dataset, as detailed in rows 2 to 6 of Table \ref{table_reconstruct}, we observed that the quality of reconstruction and the AUC of membership inference is optimal when \( \beta=0 \). Under this condition, the method can recover the highest level of detail from the original images. This is reflected in the recorded MSE for reconstruction, which is the lowest observed value at 231.31. At the same time, the inference AUC is the highest at 0.674. When $\beta=0$, the model will not distort the information of $X$ from the representation $Z$, making the model similar to normal unlearning methods (HBU and VBU) that have not considered information compression.

The reconstruction MSE rises, and the inference AUC decreases, with increasing $\beta$, aligning with our prior analysis that a larger $\beta$ results in greater distortion. Consequently, $Z$ contains less information about $X_e$. As $\beta$ increases to 1 in our experiments, the KLD between $p(Z|X)$ and $q(Z)$ attains its minimal value. This indicates that the representation $Z$ becomes most akin to the general prior, achieving an optimal level of distortion. This heightened distortion renders the reconstruction process more challenging. At such a distortion ratio, attackers are likely to reconstruct only a blurry image from $Z$, which lacks much of the detailed information present in the original images. Correspondingly, the MSE for reconstruction is at its highest at this point, registering at 668.07. Conversely, as $\beta$ increases, there is a slight decrease in prediction accuracy. Specifically, the accuracy drops from $97.45\%$ when $\beta=0$ to $95.75\%$ when $\beta=1$.

In our experiments with the Fashion MNIST dataset, detailed in rows 7 to 11 of Table \ref{table_reconstruct}, we noted trends akin to those observed in the MNIST dataset. Specifically, as $\beta$ increases, the model's learned representation increasingly resembles the assumed general prior. This similarity makes the tasks of reconstructing the original images and inferring membership more challenging. Specifically, the increase in $\beta$ from 0 to 1 leads to a rise in the difficulty of reconstruction and membership inference. The images reconstructed under higher $\beta$ values tend to lose more detailed information from the original images. Correspondingly, the MSE for reconstruction increases from 220.31 to 561.73, and the AUC decreases from 0.656 to 0.515. At the same time, the KLD to the assumed prior decreases from 12.23 to 0.13, indicating that the representation is becoming more similar to the general prior. Alongside these changes, the prediction accuracy of the model experiences a minor decrease, moving from $87.95\%$ at $\beta=0$ to $86.15\%$ at $\beta=1$.

The defense evaluations on the CIFAR10 dataset, as detailed in row 12 of Table \ref{table_reconstruct}, demonstrate the challenges of CIFAR10 image recovery. Given the dataset's complexity, unlike MNIST and Fashion MNIST, the intricate nature of CIFAR10 images makes the visual assessment of reconstruction quality less straightforward. However, we can derive conclusions similar to those from the other datasets by examining the reconstruction MSE, membership inference AUC, and model accuracy. When $\beta=0$, attackers are able to reconstruct images from CIFAR10 with an MSE of 1174.33. As $\beta$ increases, the difficulty of reconstruction also rises, with the MSE climbing to 1550.24 at $\beta =1$. This trend indicates that a larger $\beta$ leads to the discarding of more information from the model's representation, albeit at the cost of a slight decrease in prediction accuracy. Specifically, on the CIFAR10 dataset, the accuracy decreases from $84.75\%$ at $\beta=0$ to $82.95\%$ at $\beta=1$. At the same time, more information distorted hinders membership inference attacks, resulting in decreasing inference AUC from 0.821 to 0.571.

In short, CRFU can achieve a better reconstruction and membership inference attack defense when the distortion ratio $\beta$ is larger. Simultaneously, as $\beta$ increases, the accuracy drops a little. However, compared with the improvement of the defense effect on the reconstruction and membership inference attacks, we consider this accuracy reduction insignificant in the experimental parameters range.

\begin{table}[t]
	\scriptsize
	\caption{Overall Unlearning Effectiveness and Efficiency Evaluation on MNIST, Fashion MNIST, CIFAR10 and STL-10 \vspace{-2mm}}
	\label{tab_overall}
	\resizebox{\linewidth}{!}{
		\begin{tabular}{cccccc}
			\toprule
			\multirow{2}{*} { \makecell[c]{MNIST} } & \multicolumn{5}{c} {${\it EDR}=6\%$, Rep.: MLP, App.: MLP} \\
			\cmidrule(r){2-6}
			& Origin  & HBU   &VBU		  		 & CRFU  		& Retrain 		 \\
			\midrule
			Running Time (s)         & 44      	 &2.42	   & 0.20 & \textbf{0.15}                  & 41.36 	     \\
			Acc. on test dataset    & 97.6\% &87.99\% & 92.97\%      &\textbf{94.72\%} & 97.63\%    \\
			Backdoor Acc.             & 100\%  & \textbf{0.04\%}   &0.42\%         & 0.58\% & 0.08\%     \\
			\midrule
			\multirow{2}{*} { \makecell[c]{ Fashion MNIST} } & \multicolumn{5}{c} {${\it EDR}=6\%$, Rep.: MLP, App.: MLP} \\
			\cmidrule(r){2-6}
			& Origin  & HBU   &VBU		  		 & CRFU  		& Retrain 		 \\
			\midrule
			Running Time (s)         & 190.8      	 &12.67	   &1.99 & \textbf{1.56}                  & 179.35	     \\
			Acc. on test dataset    & 88.4\%      & 76.13\%&81.02\%      &\textbf{82.96\%} & 88.6\%    \\
			Backdoor Acc.             & 100\%  &  \textbf{ 0.04\%}   &1.66\%   & 1.63\% & 0.08\%     \\
			\midrule
			\multirow{2}{*} {CIFAR10} & \multicolumn{5}{c} {${\it EDR}=6\%$, Rep.: Resnet18, App.: MLP} \\
			\cmidrule(r){2-6}
			& Origin 		& HBU 		&VBU  	 		& CRFU  			& Retrain  \\
			\midrule
			Running Time (s)     & 552           & 28.21     & 0.77   & \textbf{0.72}               &518.88 \\
			Acc. on test dataset & 81.16\%	    & 74.32\%   &74.16\%     & \textbf{75.17\%}&86.65\% \\
			Backdoor Acc. & 99.83\%    &1.27\%        & \textbf{1.13\% }     & 1.7\%             &0.03\% \\
			\midrule
			\multirow{2}{*} {STL-10} & \multicolumn{5}{c} {${\it EDR}=6\%$, Rep.: Resnet18, App.: MLP} \\
			\cmidrule(r){2-6}
			& Origin 		& HBU 		&VBU  	 		& CRFU  			& Retrain  \\
			\midrule
			Running Time (s)     & 497           & 31.44     & \textbf{10.60}   & 11.24            &467 \\
			Acc. on test dataset & 63.38\%	    & 50.53\%   &50.40\%     & \textbf{52.67\%}&61.65\% \\
			Backdoor Acc. & 99.93\%    & \textbf{1.33\%}        & {1.46\% }     & 1.67\%             &0.01\% \\
			\bottomrule
	\end{tabular}}
\end{table}

\begin{figure*}[t]
	\centering  
	\subfigure[Different EDR on MNIST]{
		\begin{minipage}[t]{0.22\textwidth}
			\centering    
			\includegraphics[width=4.35125cm]{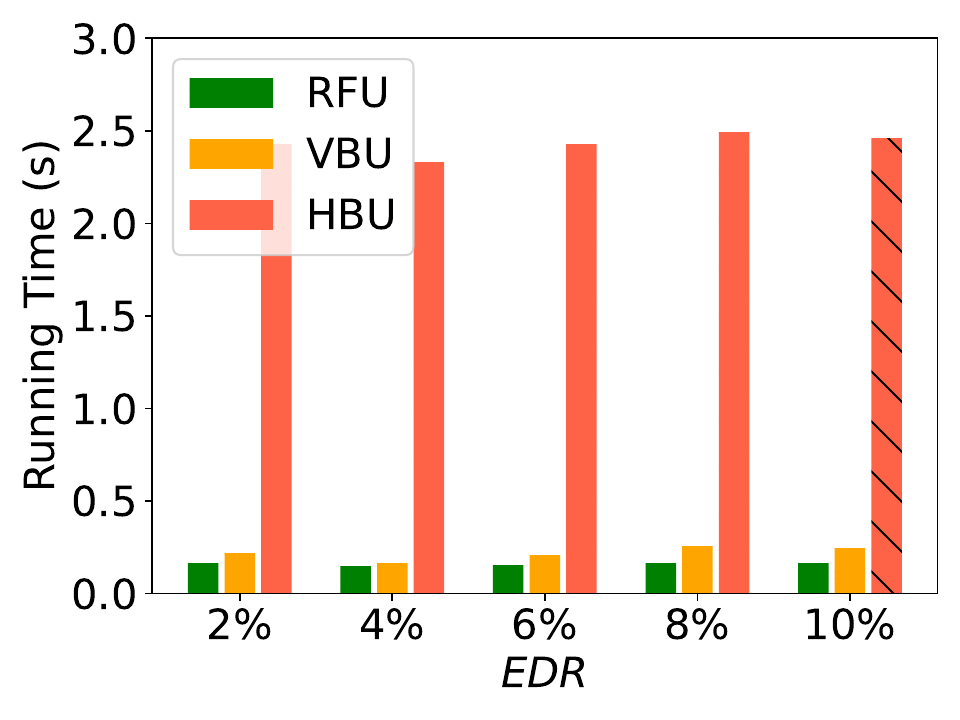}  
		\end{minipage}
		\label{mnist_rt_edr} 
	}
	\subfigure[On Fashion MNIST]{ 
		\begin{minipage}[t]{0.22\textwidth}
			\centering   
			\includegraphics[width=4.35125cm]{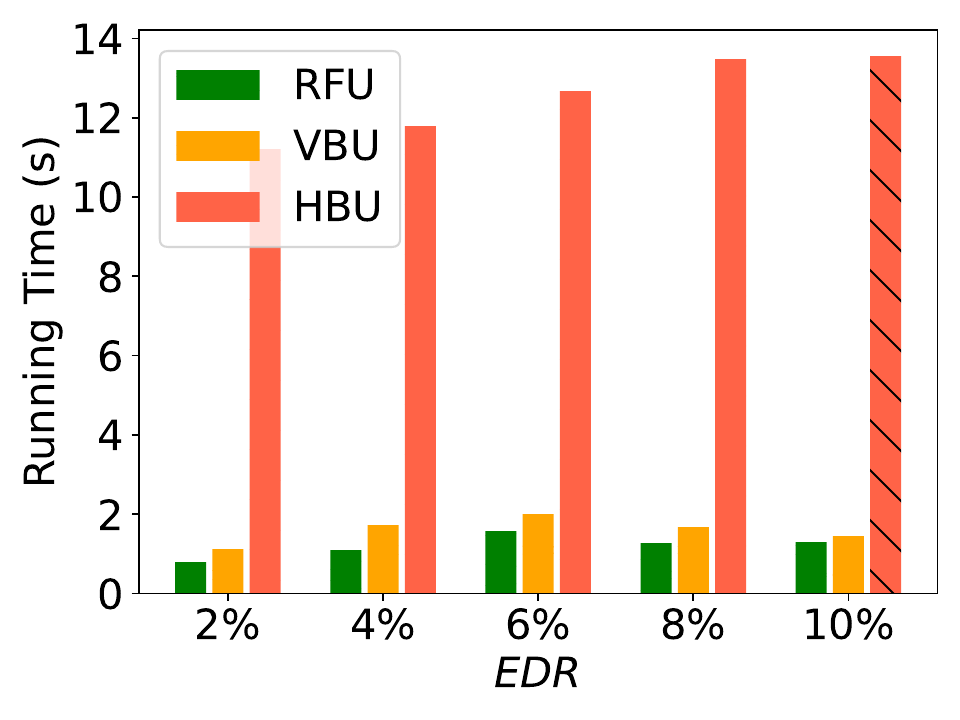}
		\end{minipage}
		\label{fmnist_rt_edr} 
	}
	\subfigure[Different EDR on CIFAR10]{ 
		\begin{minipage}[t]{0.22\textwidth}
			\centering   
			\includegraphics[width=4.35125cm]{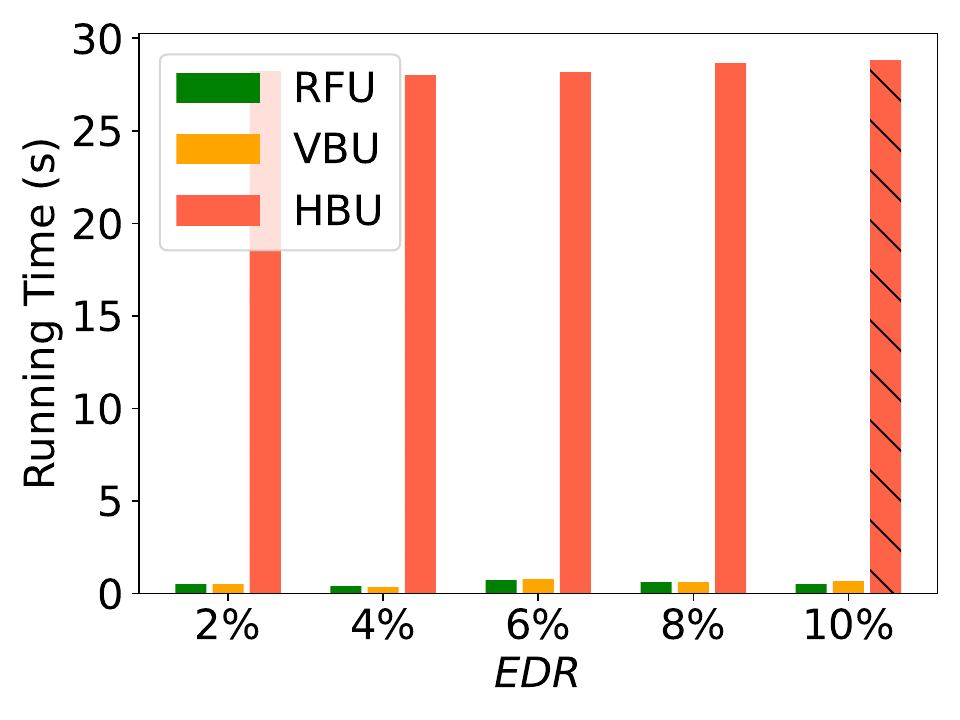}
		\end{minipage}
		\label{cifar_rt_edr} 
	}
	\subfigure[Different EDR on STL-10]{ 
		\begin{minipage}[t]{0.22\textwidth}
		\centering   
		\includegraphics[width=4.35125cm]{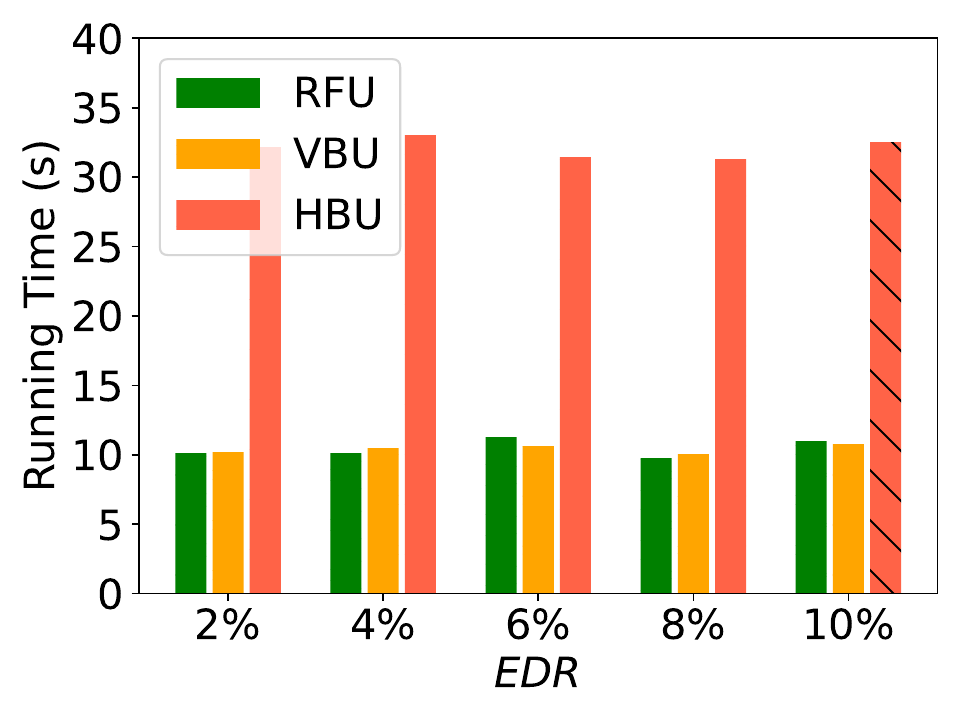}
		\end{minipage}
		\label{stl10_rt_edr} 
	}
	\\ 
	\subfigure[Different EDR on MNIST]{
		\begin{minipage}[t]{0.22\textwidth}
			\centering    
			\includegraphics[width=4.35125cm]{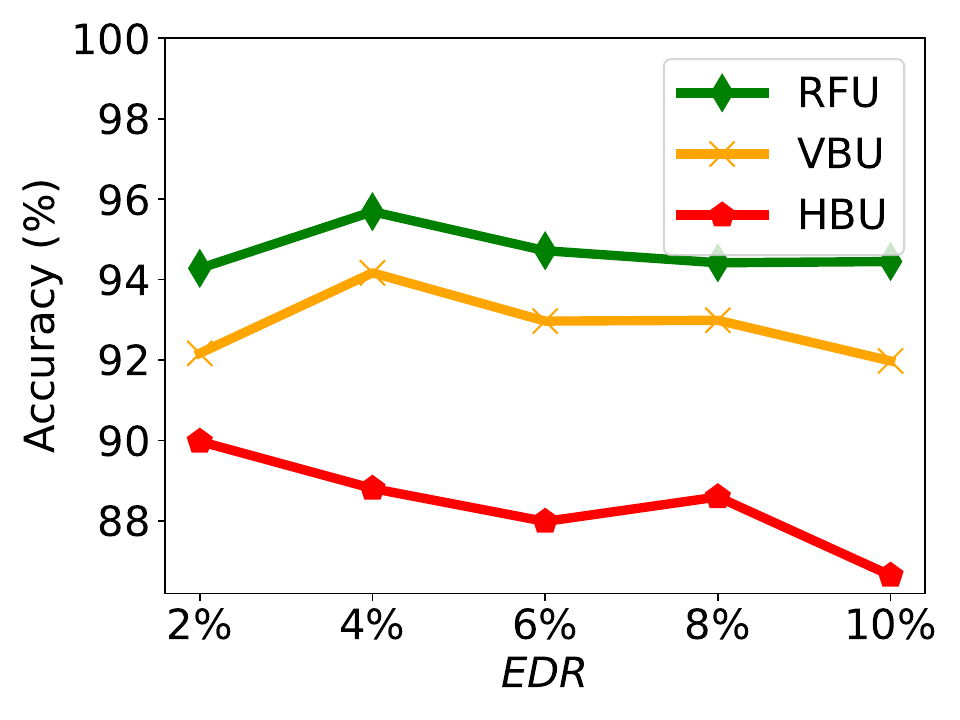} 
		\end{minipage}
		\label{mnist_acc_edr} 
	}
	\subfigure[On Fashion MNIST]{ 
		\begin{minipage}[t]{0.22\textwidth}
			\centering   
			\includegraphics[width=4.35125cm]{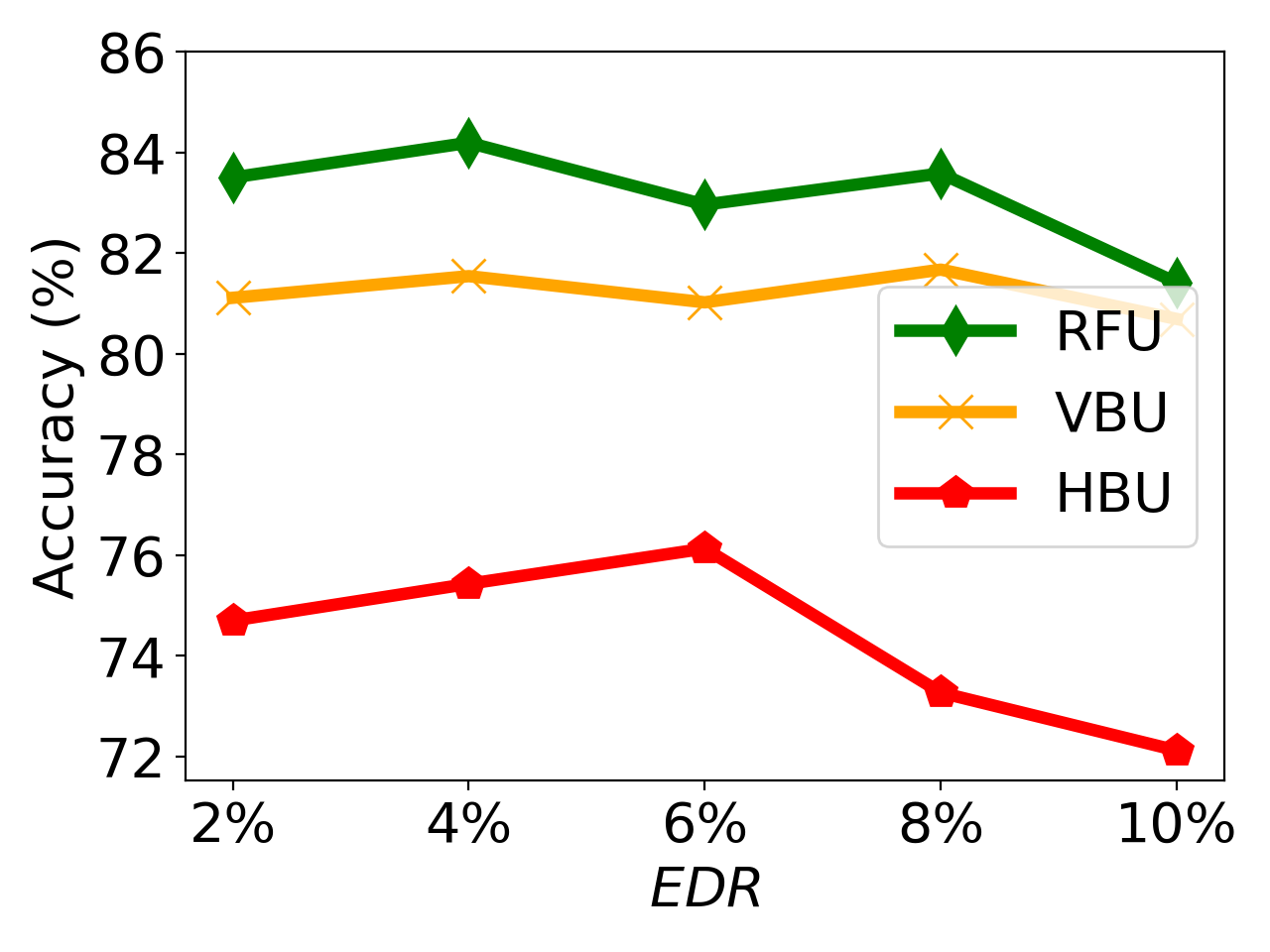}
		\end{minipage}
		\label{fmnist_acc_edr} 
	}
	\subfigure[Different EDR on CIFAR10]{ 
		\begin{minipage}[t]{0.22\textwidth}
			\centering   
			\includegraphics[width=4.35125cm]{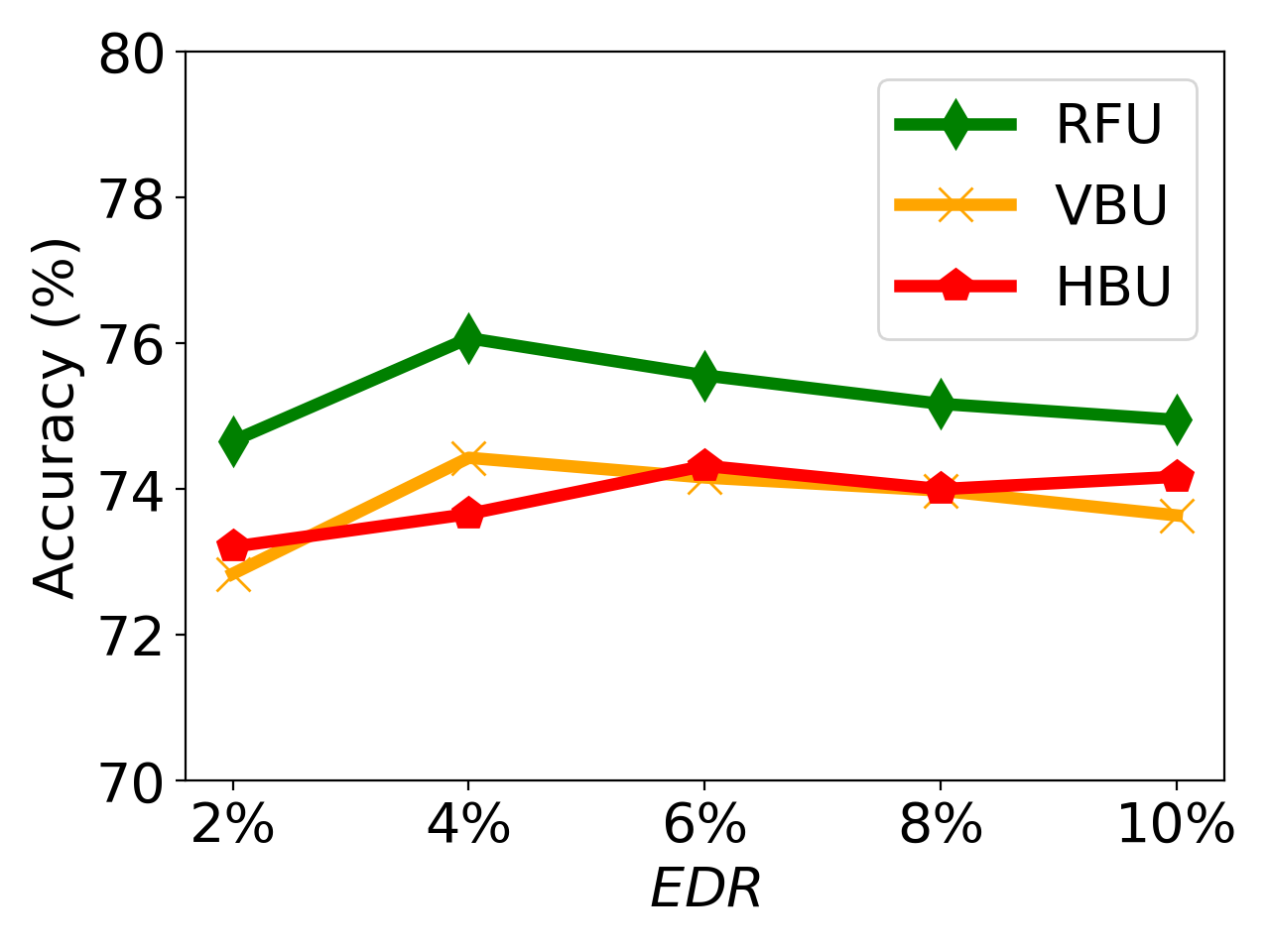}
		\end{minipage}
		\label{cifar_acc_edr} 
	}
	\subfigure[Different EDR on STL-10]{ 
		\begin{minipage}[t]{0.22\textwidth}
			\centering   
			\includegraphics[width=4.35125cm]{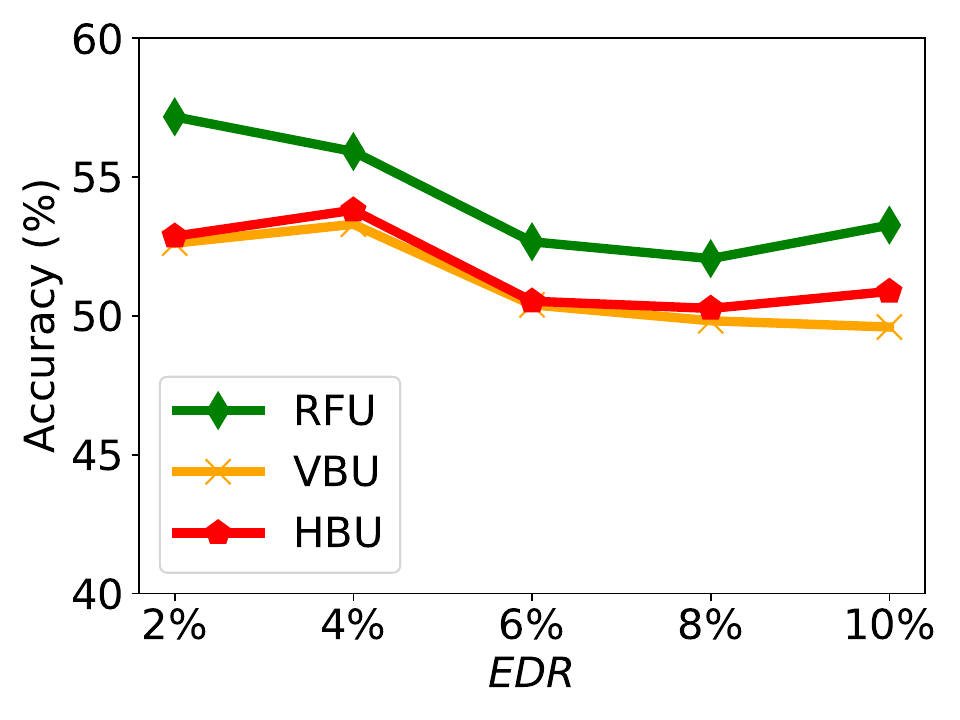}
		\end{minipage}
		\label{stl10_acc_edr} 
	}
	\\
	\subfigure[Different EDR on MNIST]{
		\begin{minipage}[t]{0.22\textwidth}
			\centering    
			\includegraphics[width=4.35125cm]{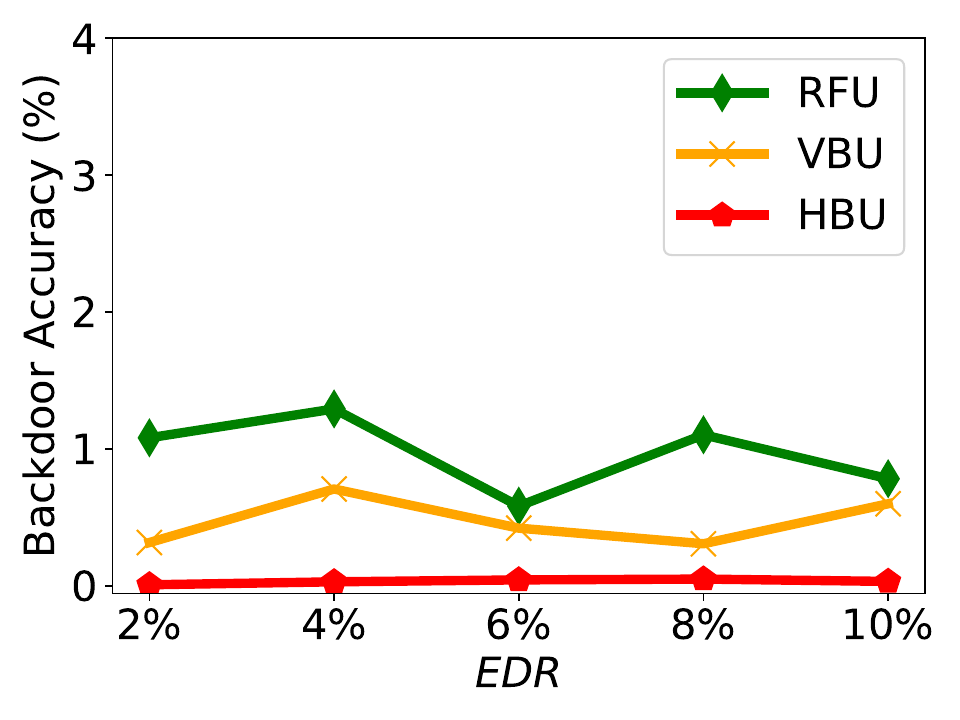}
		\end{minipage}
		\label{mnist_b_acc_edr} 
	}
	\subfigure[On Fashion MNIST]{ 
		\begin{minipage}[t]{0.22\textwidth}
			\centering   
			\includegraphics[width=4.35125cm]{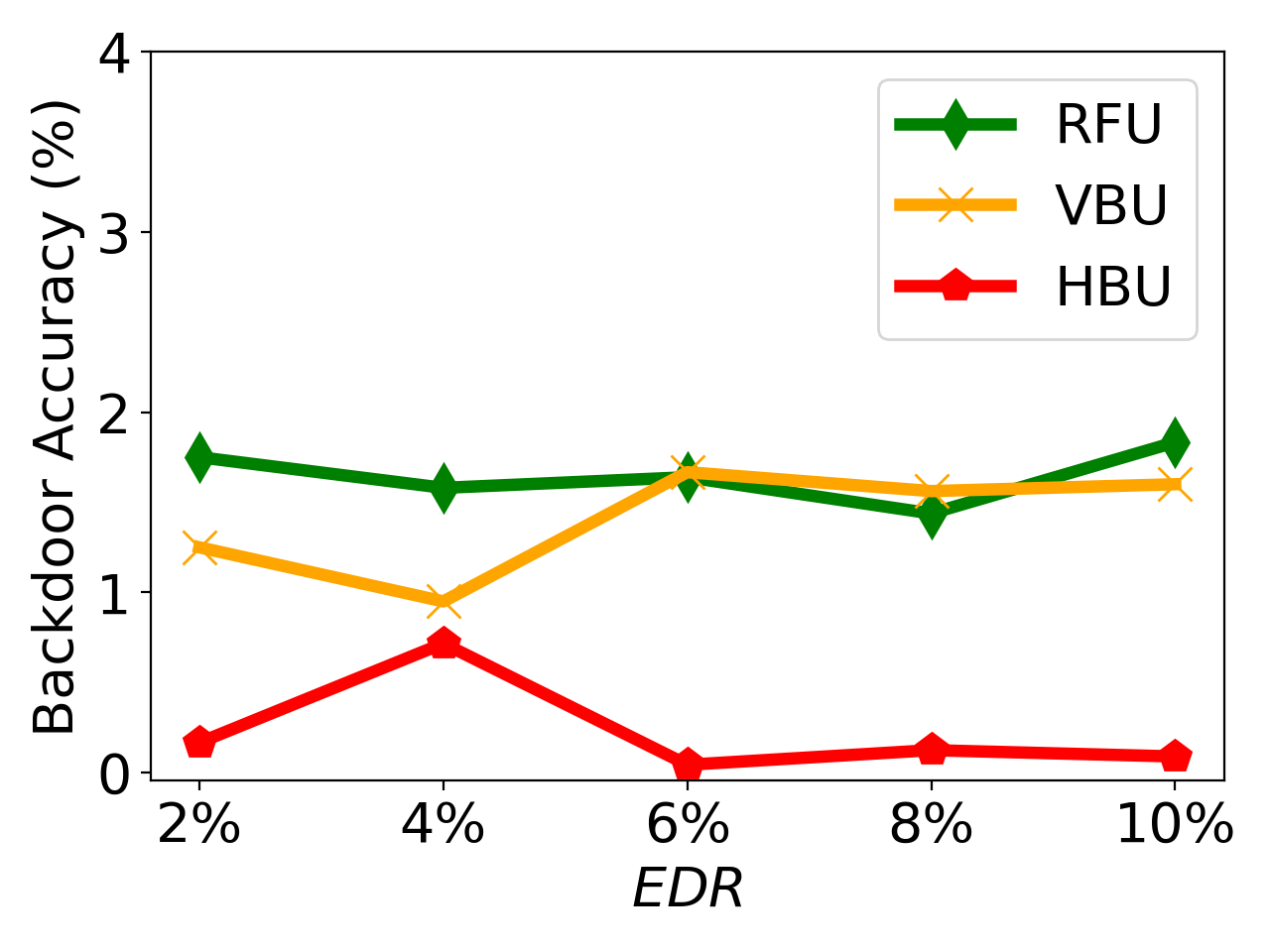}
		\end{minipage}
		\label{fmnist_b_acc_edr} 
	}
	\subfigure[Different EDR on CIFAR10]{ 
		\begin{minipage}[t]{0.22\textwidth}
			\centering   
			\includegraphics[width=4.35125cm]{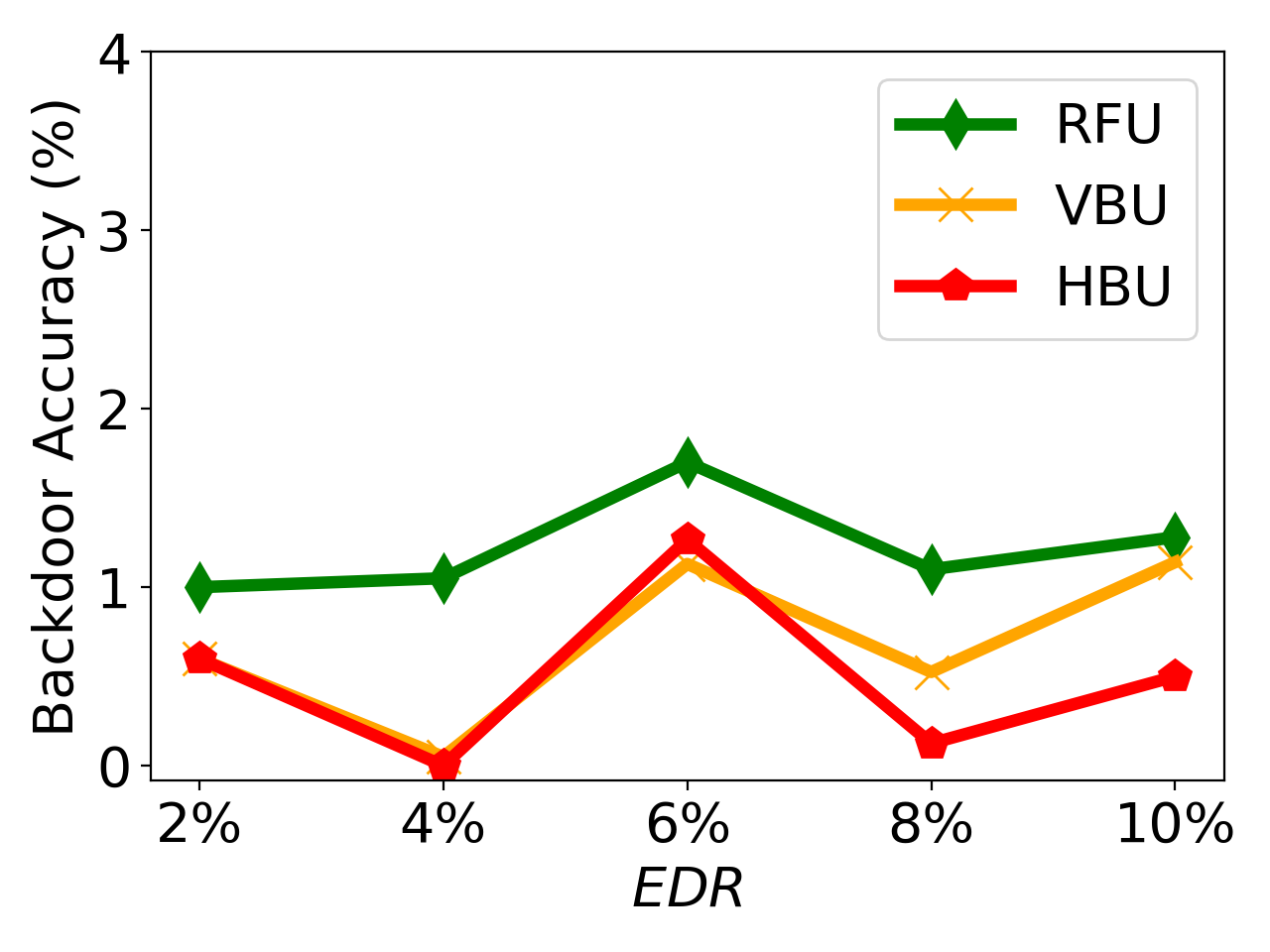}
		\end{minipage}
		\label{cifar_b_acc_edr} 
	} 
	\subfigure[Different EDR on STL-10]{ 
		\begin{minipage}[t]{0.22\textwidth}
			\centering   
			\includegraphics[width=4.35125cm]{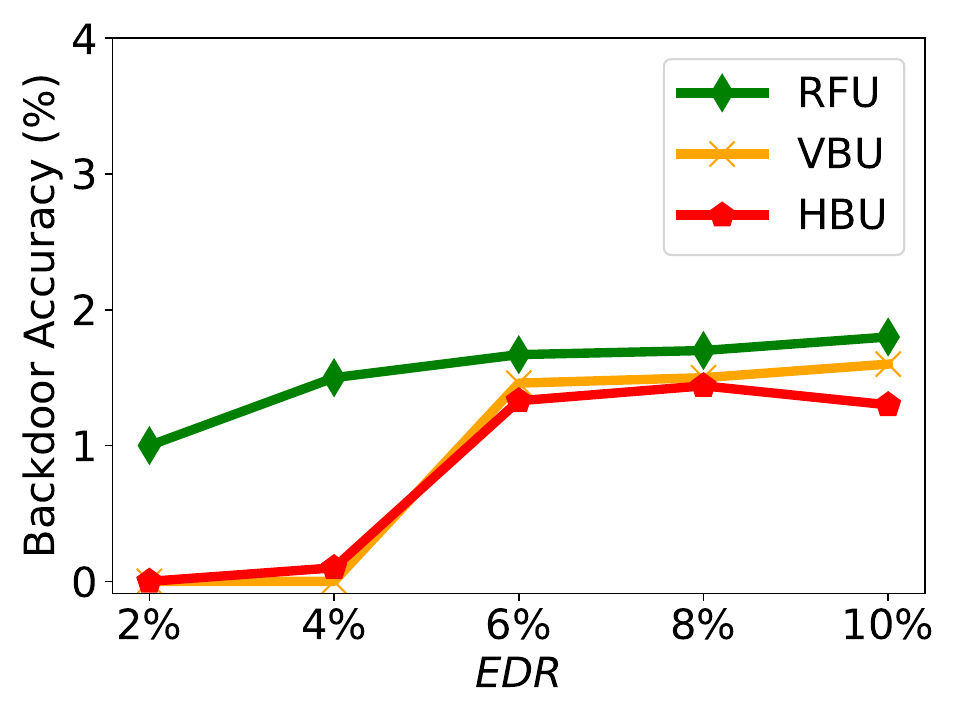}
		\end{minipage}
		\label{stl10_b_acc_edr} 
	} 
	\caption{Performance of different unlearning methods of various ${\it EDR}$}    
	\label{mnist_edr}    
\end{figure*}

\subsection{Evaluations of Model Utility}

After demonstrating the defense effect of CRFU against reconstruction attacks on unlearning, we compare the unlearning performance of CRFU and other existing state-of-the-art HBU and VBU methods. As introduced at the experiment setup, we refer to \cite{hu2022membership} to add backdoors in the erased dataset to test the unlearning effect. Although all methods successfully unlearn the erased dataset, continued unlearning training will deteriorate the model's utility. Therefore, we set a threshold of $2\%$ backdoor accuracy on the erased data, halting the model training upon reaching this threshold. This backdoor accuracy threshold guarantees the unlearning effect much better than randomly selecting, which is $10\%$ backdoor accuracy. Moreover, we set fixed $\beta$ as introduced before and set $\beta_u=0.1$ here.

We evaluate the utility of different unlearning methods from two aspects: effectiveness and efficiency. Overall results on four datasets, MNIST, Fashion MNIST, CIFAR10, and STL-10, are shown in \Cref{tab_overall}, where the erased data ratio (${\it EDR}$) is $6\%$ of the training data and unlearning rate $\beta_u=0.1$. Since these models are backdoored, accuracy on the test dataset may decrease, especially on CIFAR10, only around $81.16\%$ here. All unlearning methods successfully diminish the impact of backdoored data from the trained model lower than $2\%$ backdoor accuracy. HBU achieves the best removal effect on MNIST, Fashion MNIST, and STL-10, but it consumes the longest running time and significantly degrades the model's accuracy. 
CRFU achieves the best performance in both running time and accuracy on the test dataset most of the time. Though CRFU has not achieved the best backdoor removal effect, it effectively implements the unlearning of backdoored samples, reducing the backdoor accuracy to lower than 2\%. Detailed comparisons of different unlearning methods on MNIST, Fashion MNIST, CIFAR10, and STL-10 are demonstrated in Figure \ref{mnist_edr} and will be introduced in the following. 



\subsubsection{Efficiency of Unlearning}
We evaluate the efficiency through the running time of three unlearning methods: CRFU, VBU and HBU. 
\Cref{mnist_rt_edr,fmnist_rt_edr,cifar_rt_edr,stl10_rt_edr} show the results of the running time of all compared methods on MNIST, Fashion MNIST, CIFAR10, and STL-10. When ${\it EDR}$ is larger, it backdoors the model deeper and takes increasing time to remove the influence of these injected backdoors. It is proven on all three datasets that the running time has a slight increase as the ${\it EDR}$ increases. 
CRFU consumes a similar running time as VBU, both achieving a speedup exceeding $10\times$ when compared to HBU on MNIST, Fashion MNIST, and CIFAR10. HBU demands the most running time due to its requirement to compute the Hessian matrix using the remaining dataset to estimate the contribution of the erased data, which consumes much more time than directly unlearning based on the erased dataset. Even though HBU consumes the highest running time, it still can achieve a huge speedup compared to retraining from scratch.

\subsubsection{Effectiveness of Unlearning}
The effectiveness of unlearning is assessed based on two metrics: the model's accuracy on the test dataset and the backdoor accuracy on the erased dataset. These are presented in \Cref{mnist_acc_edr,fmnist_acc_edr,cifar_acc_edr,stl10_acc_edr} for test dataset accuracy, and \Cref{mnist_b_acc_edr,fmnist_b_acc_edr,cifar_b_acc_edr,stl10_b_acc_edr} for backdoor accuracy, respectively. First, the accuracy of the unlearned models decreases slightly as ${\it EDR}$ increases on all three datasets. The only exception is when ${\it EDR}=2\%$, unlearning this small backdoored dataset has a huge accuracy degradation than bigger ${\it EDR}$. Through extensive experimentation, we discovered that while ${\it EDR}=2\%$ backdoored samples can successfully implant a backdoor in the model, they do not embed it as deeply as a larger ${\it EDR}$ would. As a result of this shallow backdooring, unlearning the backdoor trigger using these erased samples (most features are valid) tends to damage the model more significantly. In this context, RFU achieves the best accuracy maintenance compared with the other two state-of-the-art unlearning methods, improving around $2\%$ accuracy.

Second, from the perspective of backdoor accuracy of the unlearned model, all methods effectively eliminate the impact of backdoored data, making the backdoor accuracy lower than $2\%$, which is much lower than randomly selecting $10\%$. Though HBU performs the worst in accuracy maintenance, it achieves the best backdoor removal in most of the results on the four datasets. CRFU achieves a similar backdoor removal effect as VBU on Fashion MNIST but slightly higher than VBU on CIFAR10 and STL-10.

\begin{figure}[t]
	\centering
	\includegraphics[width=0.9\linewidth]{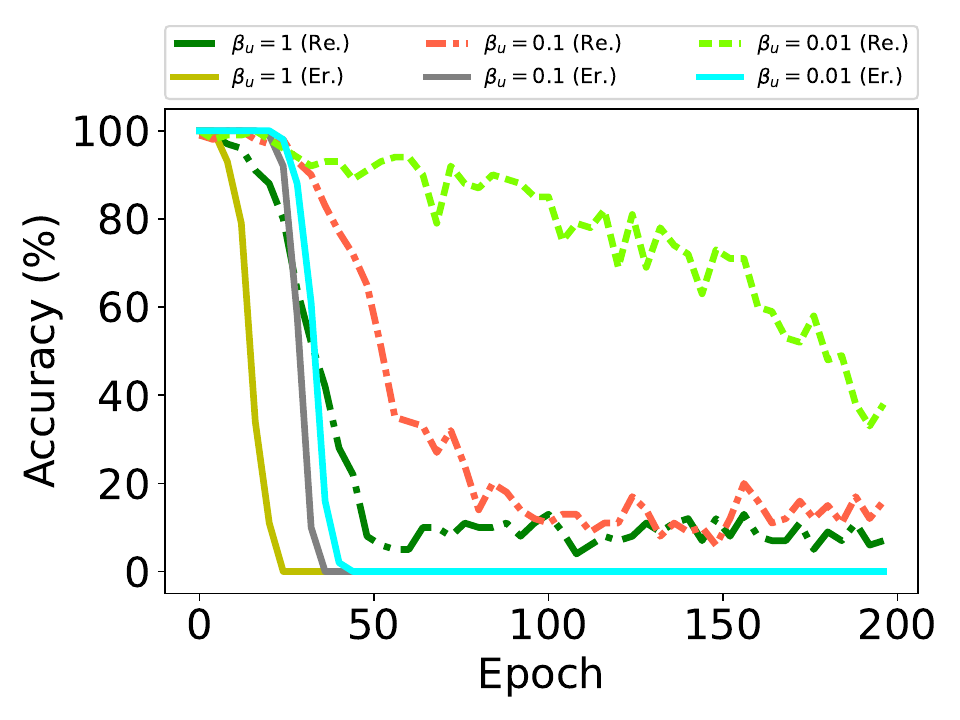}
	\caption{The variations in accuracy on the remaining (abbreviated as Re.) dataset and backdoor accuracy on the erased (abbreviated as Er.) dataset during unlearning of various unlearning rate $\beta_u$ on MNIST}
	\label{fig:mnistepochdetailacc06all}
\end{figure}

\begin{figure}[t]
	\centering
	\includegraphics[width=0.9\linewidth]{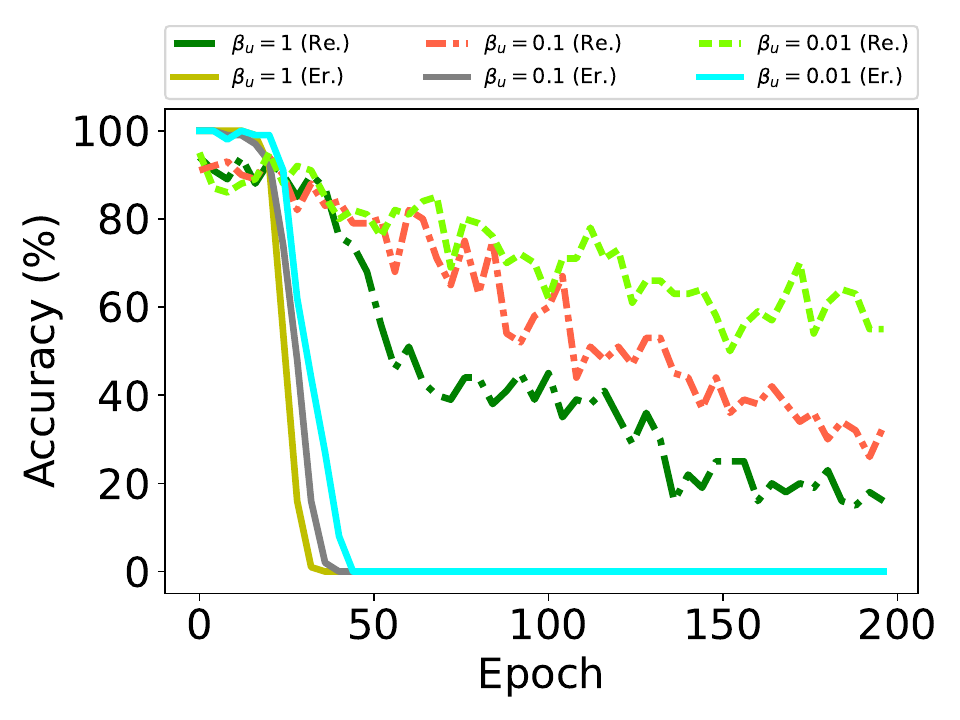}
	\caption{The variations in accuracy on the remaining (abbreviated as Re.) dataset and backdoor accuracy on the erased (abbreviated as Er.) dataset during unlearning of various $\beta_u$ on Fashion MNIST}
	\label{fig:fmnistepochdetailacc06all}
\end{figure}

\begin{figure}[t]
	\centering
	\includegraphics[width=0.9\linewidth]{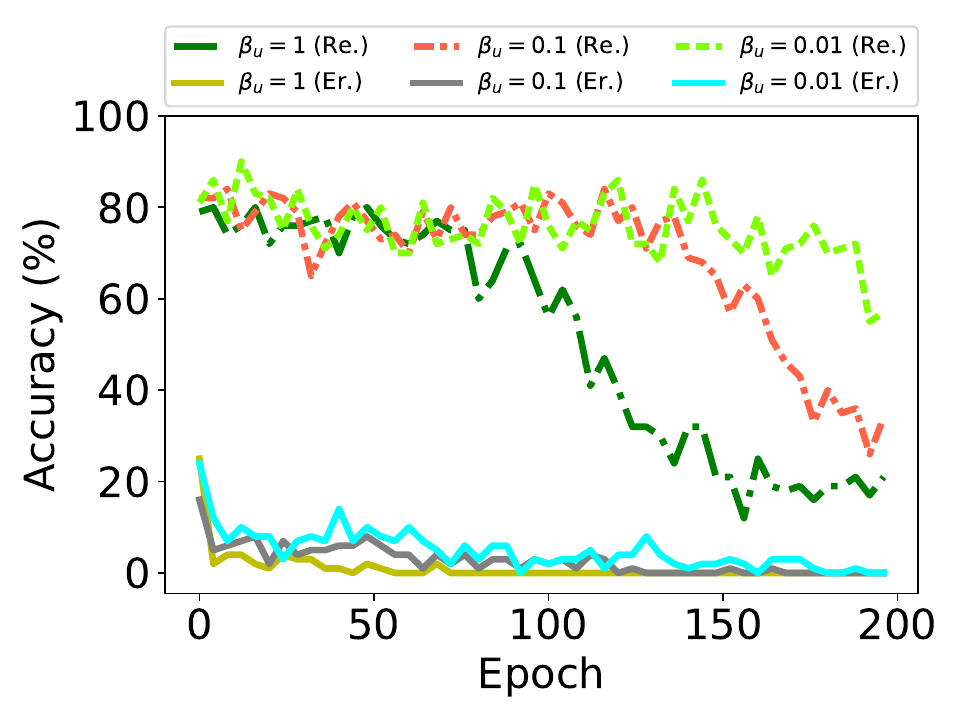}
	\caption{The variations in accuracy on the remaining (abbreviated as Re.) dataset and backdoor accuracy on the erased (abbreviated as Er.) dataset during unlearning of various unlearning rate $\beta_u$ on CIFAR10}
	\label{fig:cifarepochdetailacc06all}
\end{figure}

\subsection{Influence of the Proposed Unlearning Rate}
In this section, we evaluate the influence of the introduced unlearning rate. Since only our method has this parameter, we directly show the training progress of our method of different unlearning rate $\beta_u$ to analyze the influence of $\beta_u$ in CRFU. Specifically, we show the changes in model accuracy and backdoor accuracy in each CRFU training epoch on the remaining (abbreviated as Re.) dataset and the erased (abbreviated as Er.) dataset, respectively. To better illustrate the training process, we continue the model unlearning even when reaching the $2\%$ backdoor accuracy threshold set before. Moreover, we set $\beta=0.001$ for MNIST and Fashion MNIST and $\beta=0.0001$ for CIFAR10 and set ${\it EDR}=6\%$.

\Cref{fig:mnistepochdetailacc06all} illustrates the results of variations in accuracy and backdoor accuracy, using different $\beta_u$, from $0.01$ to $1$, on MNIST. When $\beta_u$ is large, CRFU unlearns the backdoored samples faster than $\beta_u$ is small because the larger unlearning rate $\beta_u$ means CRFU updating with a larger content on the forgetting terms of \Cref{unl_rep_p_q,app_expand}. On CIFAR10, when $\beta_u=0.01$, the accuracy of the model on the remaining dataset even does not decrease in the former 150 epochs training, but the model still removes the backdoor samples within ten epochs at the same time, as shown in \Cref{fig:cifarepochdetailacc06all}.

Controlling the unlearning speed is one of the advantages of the unlearning rate $\beta_u$; slowing down the accuracy degradation during unlearning training is another better advantage.
A slower unlearning accuracy degradation speed makes the unlearning catastrophic controllable. For example, when we choose a small $\beta_u$, such as $0.01$, when the backdoor accuracy decreases to $0$, the accuracy of the model still performs like the original model and drops slowly in the continuing training. It gives us more time to observe the unlearning process and stop the model training if accuracy degradation appears. By contrast, if we do not have the unlearning rate $\beta_u$, i.e., the $\beta_u=1$ in all situations, the catastrophic unlearning appears quickly after unlearning the erased samples, which is obvious in \Cref{fig:mnistepochdetailacc06all,fig:fmnistepochdetailacc06all}.

\section{Summary and Future Work} \label{summary}
In the paper, we propose the CRFU scheme to defend against privacy leakage attacks on machine unlearning. Specifically, we minimize the information of the erased data remaining in the learned representation to remove the contribution of specified data from the trained model. Since the representation extracts only pertinent information about labels from inputs while distorting other details, it effectively shields against privacy leakage attacks on unlearning. To avoid catastrophic unlearning, we design the remembering constraint term during data erasure and propose an unlearning rate to control the unlearning extent. Our theoretical analysis focuses on the security of CRFU in defense against reconstruction attacks. Additionally, comprehensive experimental evidence shows that our protocol can effectively counter both reconstruction and membership inference attacks based on unlearning model updates while minimizing the impact on unlearning accuracy.

Although CRFU demonstrates an effective defense against the reconstruction and membership inference attacks based on the unlearning model updates, there remains a vast uncharted area in preventing privacy leakages caused by machine unlearning. Future work should continue this line of inquiry, developing more robust privacy-preserving unlearning methods to defend against more attacks, such as adaptive attacks, where adversaries dynamically adjust their strategies based on the defense mechanisms. Additionally, as we analyzed before, the defense upper bound of CRFU is $I(Y_e;Z)$, which means adversaries can still infer the information about the unlearned labels. Exploring the gradient update forging or utilizing differential privacy techniques to overcome this challenge would be promising. Last but not least, we envision the deployment of the CRFU mechanism in practical applications, including graph analysis, point-of-interest recommendation, and medical diagnosis, thereby protecting participants' privacy.



%



\ifCLASSOPTIONcaptionsoff
\newpage
\fi



%


\small
\bibliographystyle{IEEEtranN}
\bibliography{TDSC22}

%
\begin{IEEEbiography}[{\includegraphics[width=1in,height=1.25in,clip,keepaspectratio]{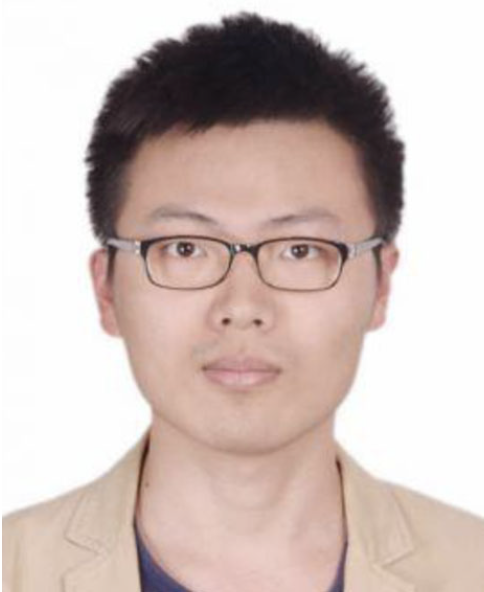}}]{Weiqi Wang}(IEEE M'24) received the M.Sc. degree in computer science from Soochow University, Suzhou, China, in 2018. He is currently pursuing the Ph.D. degree with the School of Computer Science, University of Technology Sydney, where he is advised by Prof. Shui Yu. He previously worked as a senior algorithm engineer at the Department of AI-Strategy, Local consumer services segment, Alibaba Group. He has been actively involved in the research community by serving as a reviewer for prestige journals such as ACM Computing Surveys, IEEE Communications Surveys and Tutorials, IEEE TIFS, IEEE TDSC, IEEE TIP, IEEE Transactions on SMC, and IEEE IOTJ, and international conferences such as The ACM Web Conference (WWW), ICLR, IEEE ICC, and IEEE GLOBECOM. His research interests are in machine unlearning, federated learning, and security and privacy in machine learning.
\end{IEEEbiography}
\begin{IEEEbiography}[{\includegraphics[width=1in,height=1.25in,clip,keepaspectratio]{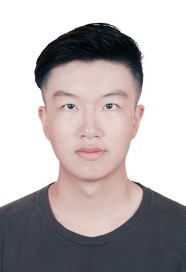}}]{Chenhan Zhang} (IEEE S'19 M'24) obtained his Ph.D. from University of Technology Sydney, Australia, in 2024, where he was advised by Prof. Shui Yu. Before that, he obtained his B.Eng. (Honours) from University of Wollongong, Australia, and  M.S. from City University of Hong Kong, Hong Kong, in 2017 and 2019, respectively. He is currently a postdoctoral research fellow at Cyber Security Hub, Macquarie University, Australia. His research interests include security and privacy in graph neural networks and trustworthy spatiotemporal cyber physical systems. He has been actively involved in the research community by serving as a reviewer for prestige venues such as ICLR, IJCAI, INFOCOM, IEEE TDSC, IEEE IoTJ, ACM Computing Survey, and IEEE Communications Surveys and Tutorials.
\end{IEEEbiography}
\begin{IEEEbiography}[{\includegraphics[width=1in,height=1.25in,clip,keepaspectratio]{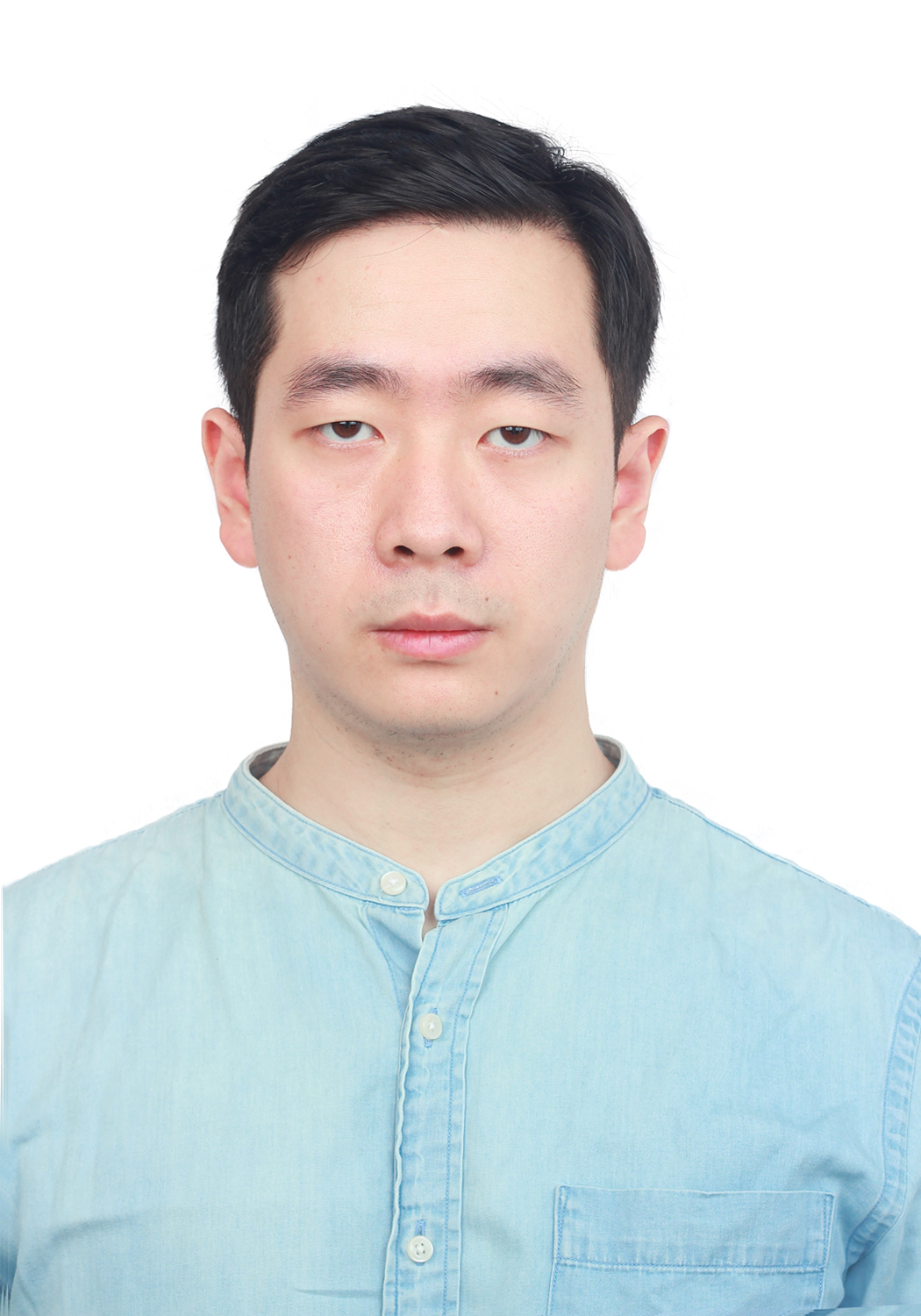}}]{Zhiyi Tian} (IEEE M'24) received the B.S. degree and the M.S. degree from Sichuan University, China, in 2017 and 2020, respectively. He received the Ph.D. degree in 2024 from University of Technology Sydney, Australia. He currently is a research associate of University of Technology Sydney, Australia. His research interests include security and privacy in deep learning, semantic communications. He has been actively involved in the research community by serving as a reviewer for prestige journals, such as ACM Computing Surveys, IEEE Communications Surveys and Tutorials, TIFS, TKDD, and international conferences, such as IEEE ICC and IEEE GLOBECOM.
\end{IEEEbiography}
\begin{IEEEbiography}[{\includegraphics[width=1in,height=1.25in,clip,keepaspectratio]{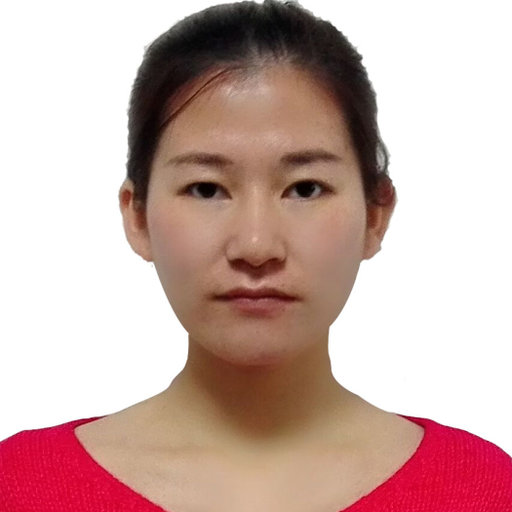}}]{Shushu Liu} (IEEE M'22)
is a researcher in Nokia bell Labs, Espoo, Finland. She received her Ph.D. degree from the Department of Communication and Networking, Aalto University, Espoo in 2023, Finland. She received the B.Sc. and M.Sc. degrees in computer science from Soochow University, Suzhou, China, in 2014 and 2017. She has been serving as the reviewer for prestigious journals such as IEEE INFOCOM, IEEE ICDCS, IEEE GlobeCom,  IEEE ICC, ACM MobileHCI, ACM SAC, etc. Her research interests include data security and privacy, 5G communication and networking, Web 3.0 and etc. 
\end{IEEEbiography}
\begin{IEEEbiography}[{\includegraphics[width=1in,height=1.25in,clip,keepaspectratio]{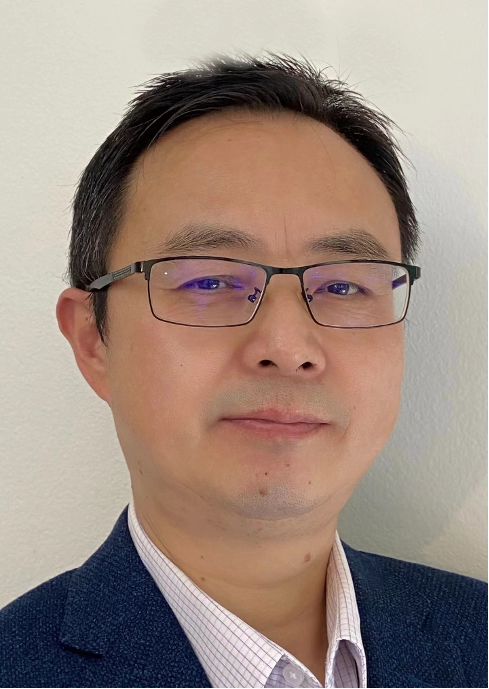}}]{Shui Yu} (IEEE F’23) obtained his PhD from Deakin University, Australia, in 2004. He is a Professor of School of Computer Science, Deputy Chair of University Research Committee, University of Technology Sydney, Australia. His research interest includes Cybersecurity, Network Science, Big Data, and Mathematical Modelling. He has published five monographs and edited two books, more than 600 technical papers at different venues, such as IEEE TDSC, TPDS, TC, TIFS, TMC, TKDE, TETC, ToN, and INFOCOM. His current h-index is 80. Professor Yu promoted the research field of networking for big data since 2013, and his research outputs have been widely adopted by industrial systems, such as Amazon cloud security. He is currently serving the editorial boards of IEEE Communications Surveys and Tutorials (Area Editor) and IEEE Internet of Things Journal (Editor). He served as a Distinguished Lecturer of IEEE Communications Society (2018-2021). He is a Distinguished Visitor of IEEE Computer Society, and an elected member of Board of Governors of IEEE VTS and ComSoc, respectively. He is a member of ACM and AAAS, and a Fellow of IEEE. 
\end{IEEEbiography}









\end{document}